\documentclass[twocolumn,twocolappendix]{aastex63}
\usepackage{xspace}
\usepackage{amsmath}

\newcommand{\angstrom}{\text{\normalfont\AA}}

\newcommand{\msunh}{\>h^{-1}\rm M_\odot}

\newcommand{\mpch}{\>h^{-1}{\rm {Mpc}}}

\usepackage{lineno}
\usepackage{color}
\received{xxx, xxx}
\revised{xxx, xxx}
\accepted{xxx}
\submitjournal{ApJ}

\shorttitle{Group/protocluster catalogs}
\shortauthors{Li et al.}
\graphicspath{{./}{figures/}}
\begin{document}

\title{Groups and protocluster candidates in the CLAUDS and HSC-SSP joint deep surveys}

\correspondingauthor{Qingyang Li, Xiaohu Yang}
\email{qingyli@sjtu.edu.cn, xyang@sjtu.edu.cn}

\author{Qingyang Li}
\affiliation{Department of Astronomy, School of Physics and Astronomy, and Shanghai Key Laboratory for Particle Physics and Cosmology, Shanghai Jiao Tong University, Shanghai 200240, China}

\author{Xiaohu Yang}
\affiliation{Department of Astronomy, School of Physics and Astronomy, and Shanghai Key Laboratory for Particle Physics and Cosmology, Shanghai Jiao Tong University, Shanghai 200240, China}
\affiliation{Tsung-Dao Lee Institute and Key Laboratory for Particle Physics, Astrophysics and Cosmology, Ministry of Education, Shanghai Jiao Tong University, Shanghai 200240, China}

\author{Chengze Liu}
\affiliation{Department of Astronomy, School of Physics and Astronomy, and Shanghai Key Laboratory for Particle Physics and Cosmology, Shanghai Jiao Tong University, Shanghai 200240, China}

\author{Yipeng Jing}
\affiliation{Department of Astronomy, School of Physics and Astronomy, and Shanghai Key Laboratory for Particle Physics and Cosmology, Shanghai Jiao Tong University, Shanghai 200240, China}
\affiliation{Tsung-Dao Lee Institute and Key Laboratory for Particle Physics, Astrophysics and Cosmology, Ministry of Education, Shanghai Jiao Tong University, Shanghai 200240, China}

\author{Min He}
\affiliation{Department of Astronomy, School of Physics and Astronomy, and Shanghai Key Laboratory for Particle Physics and Cosmology, Shanghai Jiao Tong University, Shanghai 200240, China}

\author{Jia-Sheng Huang}
\affiliation{Chinese Academy of Sciences South America Center for Astronomy (CASSACA), National Astronomical Observatories of China, Chinese Academy of Sciences, 20A Datun Road, Beijing, China}

\author{Y. Sophia Dai}
\affiliation{Chinese Academy of Sciences South America Center for Astronomy (CASSACA), National Astronomical Observatories of China, Chinese Academy of Sciences, 20A Datun Road, Beijing, China}

\author{Marcin Sawicki}
\affiliation{Department of Astronomy and Physics and Institute for Computational Astrophysics, Saint Mary’s University, 923 Robie Street, Halifax, NS, B3H 3C3, Canada}

\author{Stephane Arnouts}
\affiliation{CNRS, CNES, LAM, Aix Marseille Universit$\acute{e}$, 38 rue F. Joliot-Curie, F-13388, Marseille, France.}

\author{Stephen Gwyn}
\affiliation{NRC Herzberg Astronomy and Astrophysics, 5071 West Saanich Road, Victoria BC V9E 2E7, Canada}

\author{Thibaud Moutard}
\affiliation{CNRS, CNES, LAM, Aix Marseille Universit$\acute{e}$, 38 rue F. Joliot-Curie, F-13388, Marseille, France.}

\author{H.J. Mo}
\affiliation{Department of Astronomy, University of Massachusetts Amherst, MA 01003, USA}

\author{Kai Wang}
\affiliation{Department of Astronomy, Tsinghua University, Beijing 100084, China}
\affiliation{Department of Astronomy, University of Massachusetts Amherst, MA 01003, USA}

\author{Antonios Katsianis}
\affiliation{Department of Astronomy, School of Physics and Astronomy, and Shanghai Key Laboratory for Particle Physics and Cosmology, Shanghai Jiao Tong University, Shanghai 200240, China}

\author{Weiguang Cui}
\affiliation{Institute for Astronomy, University of Edinburgh, Royal Observatory, Edinburgh EH9 3HJ, United Kingdom}

\author{Jiaxin Han}
\affiliation{Department of Astronomy, School of Physics and Astronomy, and Shanghai Key Laboratory for Particle Physics and Cosmology, Shanghai Jiao Tong University, Shanghai 200240, China}

\author{I-Non Chiu}
\affiliation{Tsung-Dao Lee Institute and Key Laboratory for Particle Physics, Astrophysics and Cosmology, Ministry of Education, Shanghai Jiao Tong University, Shanghai 200240, China}

\author{Yizhou Gu}
\affiliation{Department of Astronomy, School of Physics and Astronomy, and Shanghai Key Laboratory for Particle Physics and Cosmology, Shanghai Jiao Tong University, Shanghai 200240, China}

\author{Haojie Xu}
\affiliation{Department of Astronomy, School of Physics and Astronomy, and Shanghai Key Laboratory for Particle Physics and Cosmology, Shanghai Jiao Tong University, Shanghai 200240, China}




\begin{abstract}
Using the extended halo-based group finder developed by \citet{Yang2021}, which is able to deal with galaxies via spectroscopic and photometric redshifts simultaneously, we construct galaxy group and candidate protocluster catalogs in a wide redshift range ($0 < z < 6$) from the joint CFHT Large Area $U$-band Deep Survey (CLAUDS) and Hyper Suprime-Cam Subaru Strategic Program (HSC-SSP) deep data set. Based on a selection of 5,607,052
galaxies with $i$-band magnitude $m_{i} < 26$ and a sky coverage of $34.41\ {\rm deg}^2$, we identify a total of 2,232,134 groups, within which 402,947 groups have at least three member galaxies. We have visually checked and discussed the general properties of those richest groups at redshift $z>2.0$. By checking the galaxy number distributions within a $5-7\mpch$  projected separation and a redshift difference $\Delta z \le 0.1$ around those richest groups at redshift $z>2$, we identified a list of 761, 343 and 43 protocluster candidates in the redshift bins $2\leq z<3$, $3\leq z<4$ and $z \geq 4$, respectively. In general, these catalogs of galaxy groups and protocluster candidates will provide useful environmental information in probing galaxy evolution along the cosmic time.
\end{abstract}



\section{Introduction} \label{sec:intro}

During the past two decades, great achievements have been made to build the galaxy-halo connections, which enabled us to better understand the galaxy formation processes, to infer the cosmological parameters and to probe the properties and distribution of dark matter \citep[see][for a recent review]{Wechsler2018}. Apart from the theoretical approaches to model the galaxy-halo connections ranging from empirical models, such as halo occupation models and conditional luminosity functions \citep[e.g.][]{Jing1998, Peacock2000, Yang2003}, to physical models such as semi-ananlytical models or hydrodynamical simulations \citep[e.g.][]{Kauffmann1993, Springel2005, Cui2012, Vogelsberger2014, Schaye2015, Cui2022}, there is also a direct way of studying the galaxy-halo connection by using galaxy groups/clusters, which are defined as sets of galaxies that reside in the same dark matter halos. 

Relatively {\it complete} galaxy group and cluster catalogs were successfully constructed from various large galaxy surveys, especially at low redshift where extensive spectroscopic data are obtained below a (shallow) limiting magnitude, e.g., the 2-degree Field Galaxy Redshift Survey  \citep[e.g.][]{Merchan2002, Eke2004, Yang2005a,  Tago2006, Einasto2007}, the Two Micron All Sky Redshift Survey \citep[e.g.][]{Crook2007, Diaz-Gimenez2015, Lu2016, Lim2017}, and most notably the Sloan Digital Sky Survey with a friends-of-friends (FOF) algorithm \citep[e.g.][]{Goto2005, Berlind2006, Merchan2005, Tempel2017}, the C4 algorithm \citep[e.g.][]{Miller2005} and the halo-based group finder developed in \citet{Yang2005a} \citep[e.g.][]{Weinmann2006, Yang2007, Yang2012, Duarte2015, Rodriguez2020}. 
Among those group finders, the halo-based group finder established in \citet{Yang2005a, Yang2007} has the particular advantage that links galaxies to their common dark matter halos \citep[e.g.][]{Campbell2015, Lu2015, Wang2020, Tinker2020}. 
Thus constructed group catalogs can be used to study the properties of galaxies as a function of their halo and group properties, and to probe how the member galaxies evolve within different environments \citep[e.g.][]{Yang2005c, Collister2005, vandenBosch2005, Robotham2006, Zandivarez2006, Weinmann2006, Wang2018}. Furthermore, these groups associated with dark matter halos can also be used to trace the large-scale structure (LSS) of the Universe \citep[e.g.][]{Yang2005b, Yang2006, Coil2006}.

On the other hand, for deeper surveys with the aim of probing galaxy properties and their evolution at high redshifts, the resulting {\it complete} group or cluster catalogs are still quite limited.  Nevertheless, group or cluster catalogs from small area surveys are obtained, e.g., from the high-redshift CNOC2 survey \citep{Carlberg1999}, DEEP2 survey \citep[][]{Gerke2005}, the zCOSMOS \citep{Wang2020}, or from photometric galaxy samples, e.g. using red-sequence cluster finders \citep{Koester2007,Rykoff2016} or other techniques \citep{Mehmood2016, Abdullah2018, Banerjee2018}. Group and cluster catalogs can also be extracted from the weak lensing \citep{Miyazaki2018}, X-ray surveys and Sunyaev–Zel’dovich (SZ) effects \citep[e.g.,][]{ACT2013,Planck2016,ACT2020,SPT2020}. In the above studies, most of the studies focus on extracting the most prominent cluster structures in the Universe and lack appropriate assessment of the overall completeness of these clusters. Very interestingly, in a recent study,  \citet{Yang2021} extended the halo-based group finder of \citet{Yang2005a, Yang2007} so that it can deal with galaxies with spectroscopic and photometric redshifts simultaneously. This new version group finder was successfully applied to the DESI image legacy surveys, where complete group catalogs ranging from low mass isolated galaxies (halos) to massive clusters in the redshift range $0<z<1.0$ with a sky coverage of 18000 square degrees were constructed.  

Due to the lack of observational data, galaxy groups and clusters at redshift beyond $z\sim 2$ are rarely studied. Most of the studies focus on the so called protocluster population \citep[see][for recent reviews]{Kravtsov2012,Overzier2016}.
The discovery of protoclusters in observations usually relies on the overdensity of star-forming regions. These regions are common at high redshift and accompanied with intrinsically high luminosity. Different sources are used to trace the star-forming areas, including H$\alpha$ emitters \citep[HAEs;][]{Cooke2014,Katsianis2017,Darvish2020,koyama2021,Shi2021}, Ly$\alpha$ emitters \citep[LAEs;][]{Venemans2007,Chiang2015,Jiang2018,Hu2021}, Lyman-break galaxies \citep[LBGs;][]{Miley2004,Toshikawa2018}, Sub-millimeter galaxies \citep[SMGs;][]{Beuther2007,Negrello2017,Cheng2019}, and the Active galactic nuclei (AGN) such as high-$z$ radio galaxies \citep[HzRGs;][]{Galametz2012,Wylezalek2013} and quasi stellar objects \citep[QSOs;][]{Capak2011}. The distribution of gas is also applied to find protoclusters \citep[e.g.,][]{Oteo2018,Miller2018}. However, the protoclusters identified with overdensity methods may not necessary to be linked with  the most massive halos at those redshifts, and thus not necessary to be able to form massive galaxy clusters at $z$=0 \citep{Cui2020}. In addition, the \emph{Planck} all sky survey provides a large sample of protocluster candidates selected by their dust emission excess in the 545 GHz band \citep{Planck2016}. The identification of protoclusters is so far challenging due to the low number density and the faintness of distant galaxies \citep{Muldrew2015}. In the past years, only a few protoclusters have been confirmed through multi-wavelength and spectroscopic analysis \citep[e.g.,][]{Diener2015,Wang2016,Lemaus2018,Polletta2021}.

From the galaxy observation side, the recently completed HSC-SSP \citep{Aihara2018} on the Subaru telescope using the HSC imager \citep{Miyazaki2018} reaches $m_i \sim 27.1$ (5$\sigma$ in 2 arcsec apertures) at Deep fields.  Although a few group catalogs using HSC data have been constructed with different methods like red-sequence galaxies \citep{Oguri2017} or weak lensing technique  \citep{Miyazaki2018,Hamana2020,Oguri2021}, and a system of galaxy protoclusters at $z\sim 4$ was searched using $g$-dropout galaxies sample selected from the Wide fields \citep{Toshikawa2018}, a few group or cluster catalogs were constructed particularly for the Deep fields \citep[e.g., an updated data version of ][]{Oguri2017}. \citet{Ando2022} searched the cores of protoclusters at $1 < z < 1.5$ using a photometric data from HSC-SSP wide and deep fields.
In addition to the $grizy$ five band photometries, the $U$-band contained in CLAUDS surveys \citep{Sawicki2019} allows bracketing the Balmer and 4000 \angstrom\ breaks at intermediate redshift that improve the performances of photometric redshift obtained from spectral energy distribution (SED) fitting \citep[e.g.,][]{Connolly1995,Sawicki1997,Sawicki2019}. 

In this study, we set out to search groups and protocluster candidates from the joint CLAUDS and HSC-SSP deep data \citep{Sawicki2019} by adopting the extended halo-based group/cluster finder developed by \citet{Yang2021}, paying particular attention to groups/protoclusters at redshift beyond $z\sim 2$. It is also worth to note that the CLAUDS and HSC-SSP deep surveys completely cover the sky areas used to study the Galaxy Evolution in the science themes of Prime Focus Spectrograph \citep[PFS;][]{Takada2014}. It would be intriguing and useful to explore the galaxy properties at high redshifts in combination with our group and protocluster candidate catalogs along with the galaxies to be observed by the PFS survey.

This paper is organized as follows. In Section \ref{sec:data}, we first introduce the data set and conditions to select galaxies, then we check the performance of the photometric redshift with respect to spectroscopic redshift and the distribution of luminosity functions for the selected galaxies sample. The extended version of halo-based group finder and the basic information of the group catalog constructed from the galaxy catalog are described in Section~\ref{sec_method}. We discuss the properties of groups at different redshifts in Section~\ref{sec_group}. In Section \ref{sec_proto}, we calculate the number density around the richest groups/clusters at different redshifts, and provide a list of protocluster candidates. Finally, we make our conclusions in Section.~\ref{sec_summary}. Throughout the paper, we adopt a $\Lambda$CDM cosmology with parameters that are consistent with the Planck 2018 results \citep{Planck2020}: $\Omega_\mathrm{m} = 0.315$, $\Omega_{\Lambda} = 0.685$, $n_{\rm s} = 0.965$, $h = H_0/(100\ \rm km\ s^{-1}\ Mpc^{-1)}=0.674$ and $\sigma_8=0.811$.

\begin{table*}
    \centering
    \caption{The selection criteria imposed on the galaxies sample using only HSC-SSP PDR2. These conditions are referring \citet{Oguri2017} and \citet{Mandelbaum2018}. The flags marked * are only True for the E-COSMOS field.}
    \label{tab:hscsel}
    \begin{tabular}{l c l}
    \hline \hline
     Conditions & True or False & Descriptions\\
     \hline
    %  \multicolumn{3}{c}{Basic cuts}\\
    % \hline
    % Mask & False & object at mask region\\
    % Obj\_type & Galaxy & object type\\
    % $m_{\rm g,r,i,z,y} > $ 0 & & apparent magnitude\\
    % $z_{\rm pho} > 0$ & & photometric redshift\\
    % $\sigma_i < $ 1 & & error of apparent magnitude\\
    % $m_{\rm i} < $ 26 & & apparent magnitude cut \\     
    %  \hline
    %  \multicolumn{3}{c}{HSC sources cuts}\\ 
    %  \hline
     \emph{isprimary} & True & Identify a single version of astrophysical object\\
     \emph{$g|r|i|z|y\_$inputcount$\_$value} $\ge2$& False & Number of images contributing at center\\
     \emph{$g|r|i|z|y\_$mask$\_$pdr2$\_$bright$\_$objectcenter} & False & Source center is close to bright object pixels\\
     \emph{$i\_$extendedness$\_$value} = 1 & True & Extended object\\
     \emph{$g|r|i|z|y\_$pixelflags$\_$edge} & False & Too close to an image boundary\\
     \emph{$g|r|i|z|y\_$pixelflags$\_$interpolatedcenter}* & False  & Interpolated pixel in the source center\\
     \emph{$g|r|i|z|y\_$pixelflags$\_$saturatedcenter} & False & Saturated pixel in the source center\\
     \emph{$g|r|i|z|y\_$pixelflags$\_$crcenter}* & False  & Cosmic ray in the source center\\
     \emph{$g|r|i|z|y\_$pixelflags$\_$bad} & False & Bad pixel in the source center\\
    \hline
\end{tabular}
\end{table*}

\section{Galaxy samples} \label{sec:data}

In this section, we describe the data sets used in this study and the criteria to select galaxy samples. We assess the performance of the photometric redshifts and measure the galaxy luminosity functions to evaluate our galaxy samples. 

\subsection{The photometric surveys}

We use the joint CLAUDS and HSC-SSP data set \citep{Sawicki2019}, which has been applied for studies including the UV and $U$-band luminosity functions \citep{Moutard2020} and source classification \citep{Golob2021}. This joint data set is a SExtractor-based multiband catalog as described in \citet{Sawicki2019}. The detection to an object uses the signal-to-noise ($\Sigma$SNR) image, which is constructed from all available CLAUDS $u/u^*$ and HSC-SSP $grizy$ images. Once objects are detected by the SExtractor software \citep{Bertin1996} in the $\Sigma$SNR image, the multiband catalog is then created by running SExtractor in dual image mode, with various measurement recorded for each object, such as positions, fluxes (in Kron, isophotal, and fixed-radius circular apertures), fiducial radii and ellipticities. Here, we only give a brief description of the combination of the two data sets, while more details can be found in \citet{Sawicki2019} and \citet{Moutard2020}. 

The CLAUDS and HSC-SSP data set contains $U+grizy$ six bands data, distributed in four roughly equal-sized ($\sim$4-6 $\deg^2$) fields: E-COSMOS, XMM-LSS, ELAIS-N1 and DEEP2-3. CLAUDS provides the $U$-band data with a median depth of $U_{\rm AB}=27.1$ ($5\sigma$ in 2 arcsec apertures) covering a total 18.60 $\rm deg^2$ in HSC-SSP Deep layer, and 1.36 $\rm deg^2$ sub-area reaching a depth of $U_{\rm AB}=27.7$ within UltraDeep layer \citep{Sawicki2019}. CLAUDS uses two $U$-band filters: the new $u$ filter is applied in the ELAIS-N2 and DEEP2-3 fields, while the older $u^*$ filter is adopted in XMM-LSS. The E-COSMOS field uses both $u$ and $u^*$ filters in the central region and only the $u$ filter in other areas. The median seeing in the entire deep fields of CLAUDS at $U$-band is 0.92 arcsec.  

The HSC-SSP data contains $grizy$ five wavebands with the depths of $g_{\rm AB} \sim 27.3$, $r_{\rm AB} \sim 26.9$, $i_{\rm AB} \sim 26.7$, $z_{\rm AB} \sim 26.3$ and $y_{\rm AB} \sim 25.3$ ($5\sigma$ in 2 arcsec apertures) in the Deep and UltraDeep regions \citep{Aihara2019}, respectively. The average seeing in the $i$-band is the best among the five wavebands, reaching $\sim 0.62$ arcsec. The HSC-SSP data set totally has 14,789,205 objects over 34.41 $\rm deg^2$, within which 18.60 $\rm deg^2$ have CLAUDS $U$-band observations. We note that our data set uses an updated version of HSC-SSP sample, which is based on the second public data release \citep[PDR2;][]{Aihara2019}.  
% while the analysis in \citet{Sawicki2019} used the S16A internal HSC-SSP data release.
This data version increases the galaxy number but not significantly in Deep and UltraDeep fields. In the following, we describe the procedures used to select our galaxy samples from these data sets.

\begin{figure*}
    \centering
    \includegraphics[width=0.9\textwidth]{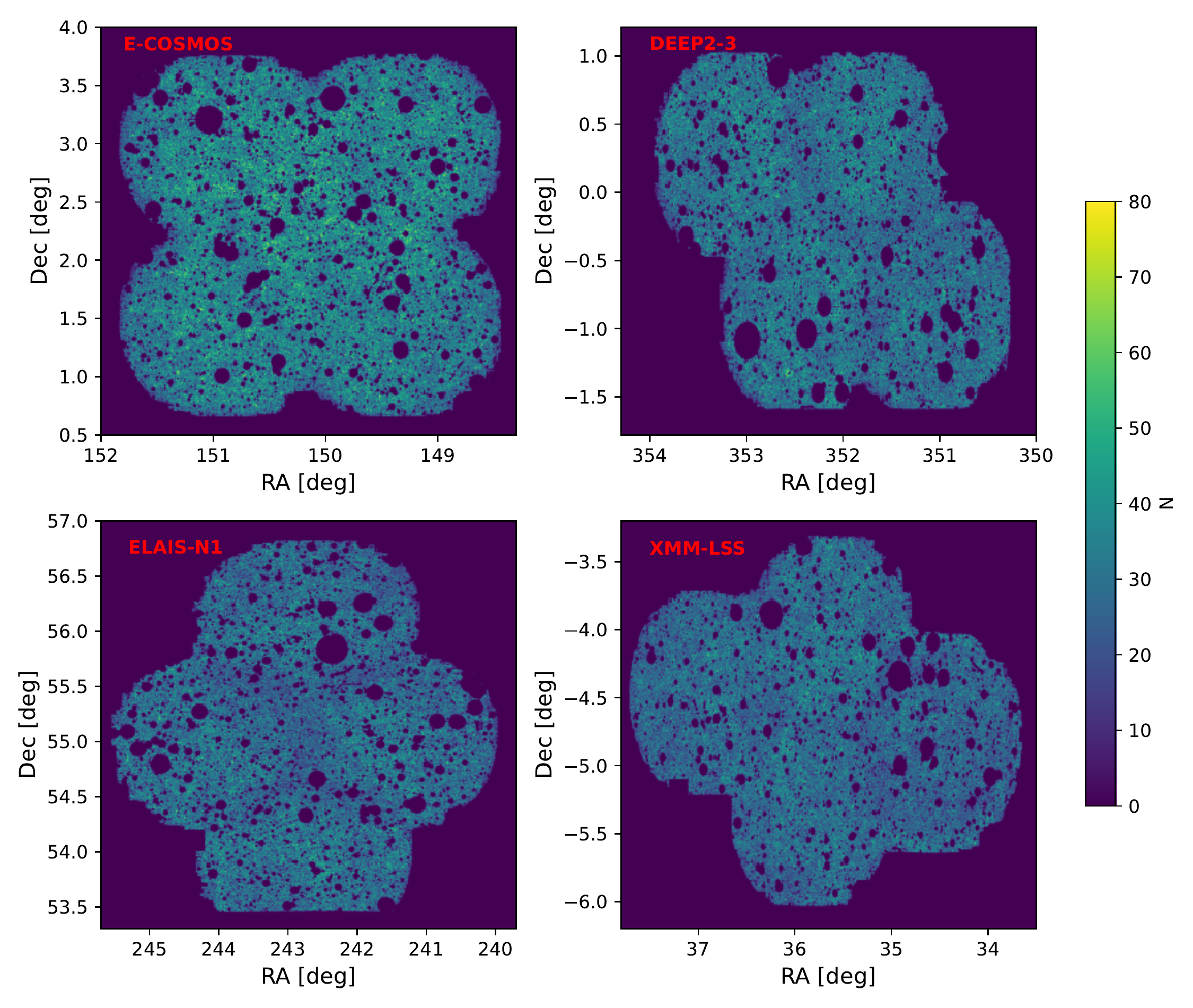}
    \caption{The sky coverage of galaxies, distributed in four separated fields: E-COSMOS, DEEP2-3, ELAIS-N1 and XMM-LSS. The galaxy number count in each pixel with an area about $1.4 \times 10^{-4}\ \rm deg^2 $ is coded with a color bar. The empty circles inside galaxies coverage correspond to the masked areas.}
    \label{fig:dis2D}
\end{figure*}

\subsection{Galaxy selection}

Our galaxy sample selection begins with the rejection of the objects with mask flags to avoid the light influence from bright stars. While the objects in the data set have been divided into stars, galaxies and QSOs using the gradient boosted trees method \citep{Golob2021}, we only select the objects classified as galaxies. We also exclude galaxies with $grizy$ apparent magnitude less than 0 to avoid spurious photometric redshift determinations. As the $i$ band is priority observed and has the smallest seeing among the HSC-SSP observations, we choose to use galaxies observed in $i$-band with apparent magnitude limit $m_i<26$ and magnitude error $\sigma_i < 1$. We discard away two small but bad areas at the edges of ELAIS-N1 and XMM-LSS. With these initial cuts for the CLAUDS and HSC-SSP data set, we have 7,752,546 galaxies spanning an area of 33.97 $\rm deg^2$.

Then, we try to exclude the unreliable galaxies due to the pollution of the bad pixels or the problems still existing in the image processing. However, the original data set provided in  \citet{Sawicki2019} does not give such information about these galaxies. We thus construct a new independent galaxy sample based on the HSC-SSP PDR2 database and apply several pixel flags of images to exclude these unreliable sources. The chosen flags refer to the criteria adopted in \citet{Oguri2017} and \citet{Mandelbaum2018}. We first require the flag \emph{isprimary} to be $True$ to identify a single version of each astrophysical object. We throw out the sources close to the bright object pixels at $grizy$ bands. The object type is determined as galaxy by setting the star-galaxy separation parameter \emph{i\_extendedness\_value=\rm 1} at $i$ band. The number of visits to each object is indicated with the \emph{inputcount} parameter, where we choose \emph{inputcount} $\geq$ 2 for all $grizy$ bands. Moreover, we also discard possibly polluted galaxies with these situations: objects too close to an image boundary or those that have interpolated, saturated, bad and a cosmic hit pixel in the source center at any broad bands. Because the ultra-deep fields undergo 100 or more visits, any one of which could be affected by a cosmic ray, resulting in a substantial chance of the object being excluded, i.e., lower source density in the ultra-deep fields than that in the deep fields after excluding the influenced sources. We thus do not apply the interpolated and cosmic ray flags for the E-COSMOS field which suffers this effect heavily, though the effects of cosmic rays are minor on coadds \citep{Aihara2021}. After applying these selection criteria as listed in Table.~\ref{tab:hscsel}, we obtain a galaxy checking sample with a total of 11,177,216 sources.

\begin{figure}
    \centering
    \includegraphics[width=0.45\textwidth]{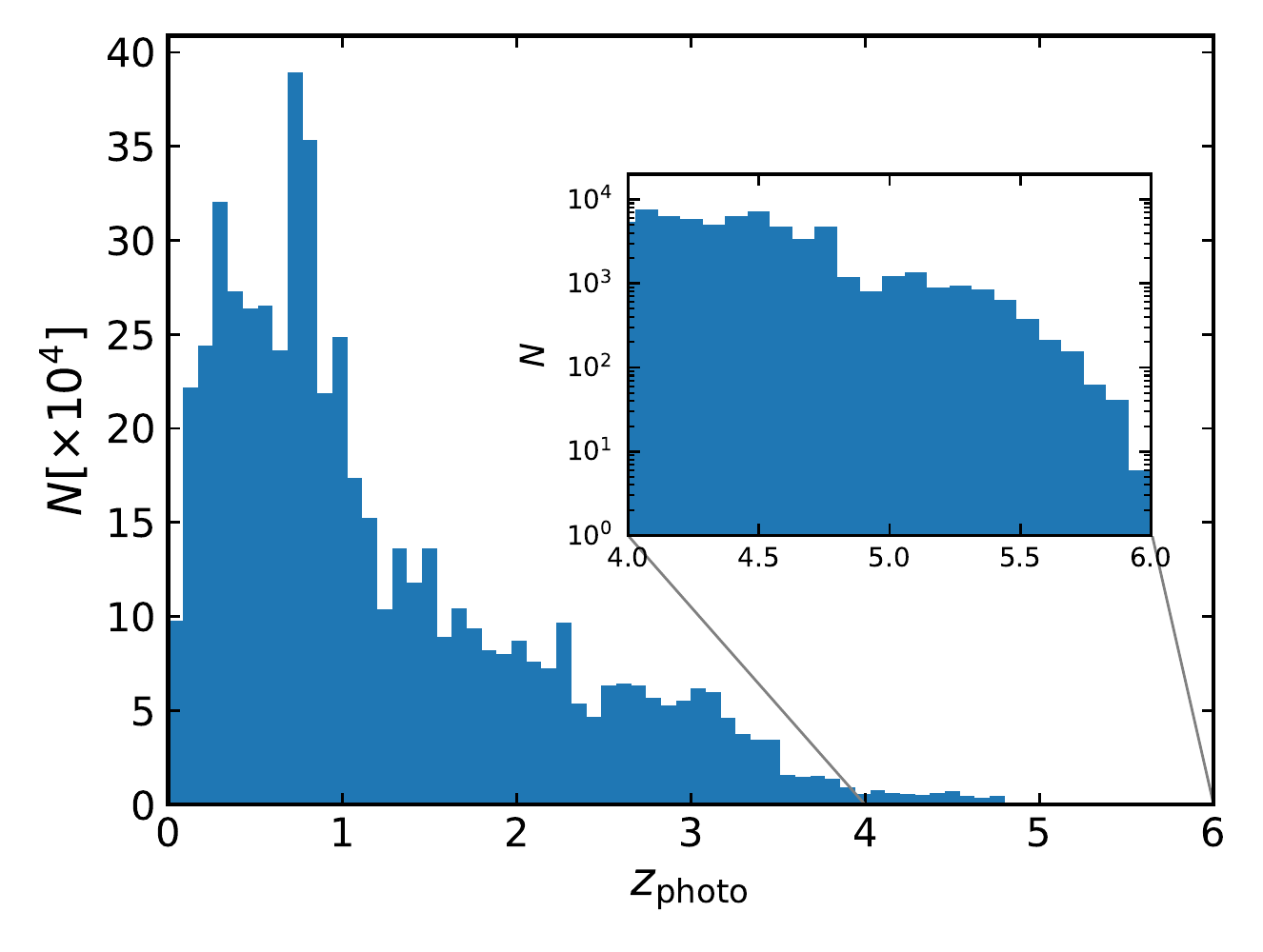}
    \caption{The number distribution of galaxies as a function of photometric redshift in our final sample. The inserted panel particularly shows the number distribution of galaxies at $4 < z_{\rm photo} < 6$.}
    \label{fig:galz}
\end{figure}

Finally, we match the initially selected CLAUDS and HSC-SSP galaxy sample with this checking sample obtained from HSC-SSP PDR2 by asking the agreement of their coordinates to be less than 1 arcsec. By selecting galaxies with photometric redshift $0<z_{\rm photo}<6$, we finally obtain a galaxy catalog with 5,607,052 galaxies. The distributions of the matched galaxies in the final sample at four separated fields are shown in different panels as quoted in Fig.~\ref{fig:dis2D}. We present the redshift distribution of selected galaxies in Fig.~\ref{fig:galz}. 

\begin{figure*}
    \centering
    \includegraphics[width=0.8\textwidth]{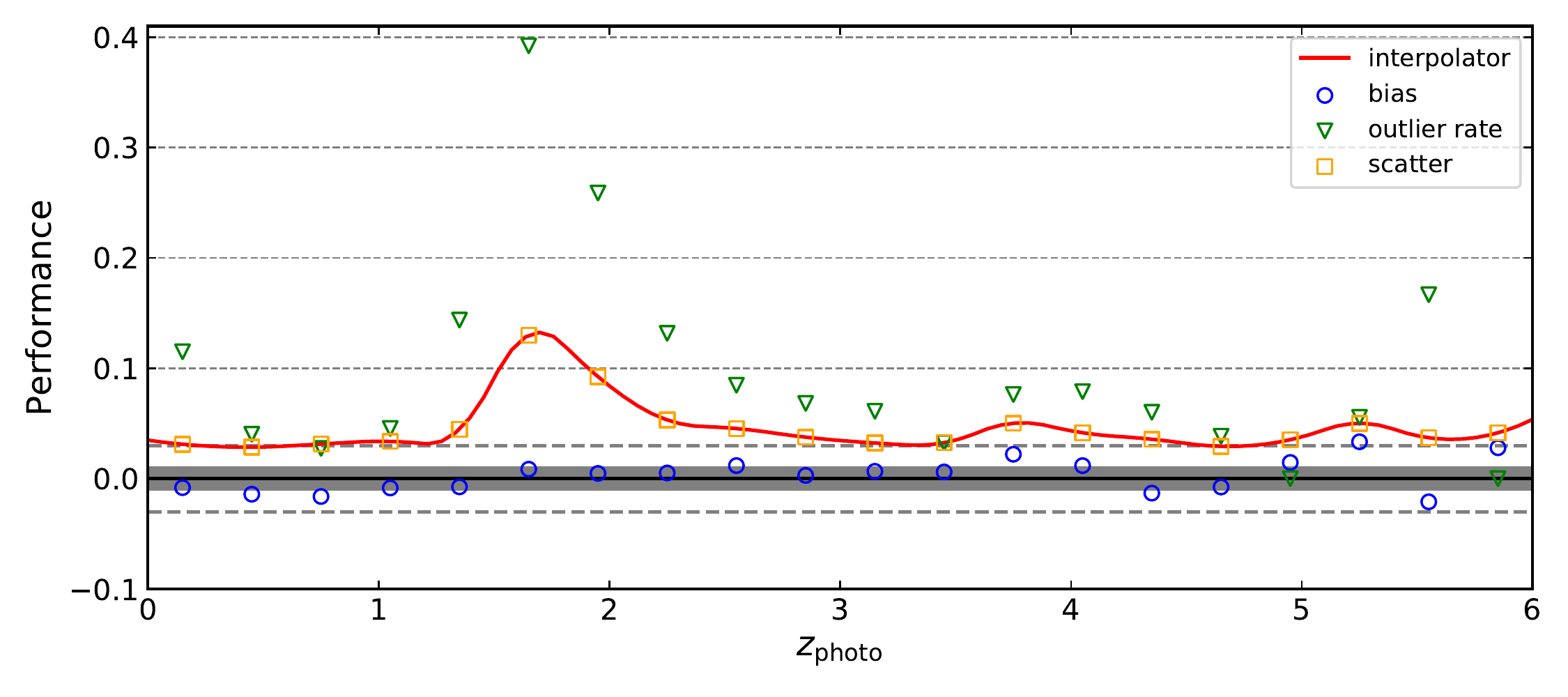}
    \caption{Assessing the quality of photometric redshifts  with respect to spectroscopic redshifts as a function of photometric redshift. The blue circles, green triangles and orange squares represent the bias, scatter and outlier rate, respectively. The grey shadows and dashed lines show $\pm0.01$ and $\pm0.03$ region, respectively. The red line is the second order spline interpolation for the scatter parameter.}
    \label{fig:redz}
\end{figure*}

\subsection{Assess the photometic redshift quality} \label{sec:zphoto}

The photometric redshift, $z_{\rm photo}$, in our galaxy sample is calculated using a color-space nearest-neighbour machine learning technique \citep[hereafter kNN;][]{Sawicki2019} combined with the template-fitting code \textsc{LE PHARE} \citep{Arnouts2002,Ilbert2006}. 
Specifically, the kNN method uses the 30-band COSMOS photometric redshifts from \citet{Laigle2016} as a training set. 50 nearest neighbours around each object are determined by the kNN in color space. Then, each object was fitted with a weighted Gaussian kernel density estimator with the weighted redshifts of these neighbours. This method obtains redshifts with a low scatter and bias on average but suffer from more outliers. The photometric redshifts from \textsc{LE PHARE} are computed by using the template library of \citet{Coupon2015} and with a consideration of four extinction laws as described in \citet{Ilbert2006}. The final photometric redshifts were obtained by combing the outputs from the kNN method and \textsc{LE PHARE}, i.e., the photometric redshift values of outliers in the kNN photo-$z$ catalog were replaced with the values from \textsc{LE PHARE}.
The details of computation are introduced in \citet{Moutard2020}. 

As the group finder uses both the value and error of photometric redshifts (see Section \ref{sec:method} for details), we investigate the quality of photometric redshifts by comparing with the subsample of high-quality spectroscopic redshifts, $z_{\rm spec}$. We use the quantities including bias, scatter and outlier rate to assess the quality of photometric redshifts. The bias is defined as the median value of $(z_{\rm photo}-z_{\rm spec})/(1+z_{\rm spec})$. The scatter is estimated using the normalized median absolute deviation: median $(|z_{\rm photo}-z_{\rm spec}|/(1+z_{\rm spec}))$/0.6745. The outlier rate is the fraction of galaxies with $|z_{\rm photo}-z_{\rm spec}|/(1+z_{\rm spec}) > 0.15$ in each photometric redshift bin. 

Based on the spectroscopic redshifts of 65135 galaxies from a compilation of surveys \citep{Lilly2007,Bradshaw2013,LeFevre2013,McLure2013,Comparat2015,Kriek2015,Silverman2015,Masters2017,Tasca2017,Scodeggio2018} in the CLAUDS and HSC-SSP sample, we assess the photometric redshift quality in our galaxy sample. The performance as a function of photometric redshifts is shown in Fig.~\ref{fig:redz}. The bias is constrained around $\pm0.01$ at $z_{\rm photo} < 3.5$ and within 0.03 at higher redshifts. The scatter remains at 3 per cent level across whole redshift range, except reaching $\sim 12$ per cent at $z\sim1.8$ where the outlier rate is as high as about 40$\%$. Thus, in general the groups we extracted at redshift  $z \sim 1.8$ should be less reliable than other redshift ranges. 
We use a second-order spline interpolation method to fit the scatter as a function of redshift, which is shown as the red line in Fig.~\ref{fig:redz}. We use this fitting result to describe the photometric redshift error of each galaxy for our group finder. If a galaxy has a spectroscopic redshift, we replace the photometric redshift with the spectroscopic redshift and set the redshift error as 0.0001. As tested in \citet{Yang2021}, galaxy groups, especially massive ones, can be reliably detected for galaxy samples with a photometric redshift error at 3\% level. We note that once the  
PFS starts its operation, we will keep updating the photometric redshifts with spectroscopic redshifts, and hence improve the resulting group catalogs. We expect the completeness and purity, and more importantly, the redshift accuracy of  of the groups in the updated versions will be improved.

\begin{figure*}
    \centering
    \includegraphics[width=0.9\textwidth]{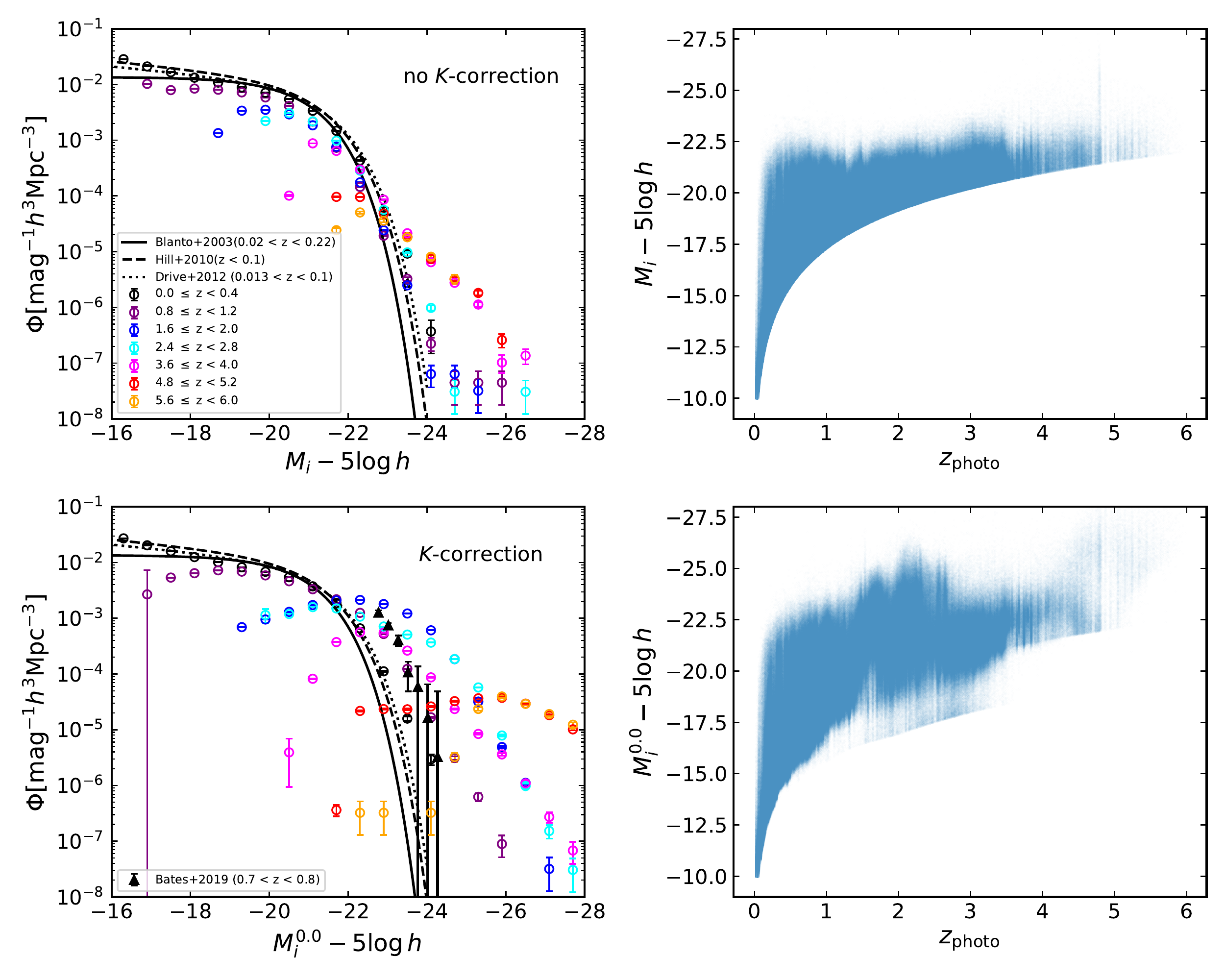}
    \caption{Left panels: galaxy luminosity functions in different redshift bins. The results for galaxies with or without $K$-corrections are shown in the lower and upper panels, respectively. The errors are derived from the square root of the sum of weights squared (Possion error). Seven redshift bins with 0.4 interval are shown with different colors as labelled in the legend in the top panel. The comparison with other observations at low redshifts in the $i$-band include \citet{Blanton2003} (solid black), \citet{Hill2010} (dashed black) and \citet{Driver2012} (dotted black). The black triangles show a relatively higher-redshift observation result with $K$-correction in \citet{Bates2019}. Right panels: the absolute magnitude v.s. photometric redshift distribution of a subset of randomly selected sample galaxies.
    The results for absolute magnitudes with  or without  $K$-corrections are shown in the lower and upper panels, respectively.  }
    \label{fig:lum_gal}
\end{figure*}

\subsection{Galaxy luminosity functions}

As the halo-based group finder uses the group/galaxy luminosity as a proxy for halo mass estimates, it is important to check if there are spuriously bright galaxies due to redshift errors etc. 
The absolute magnitude of each galaxy is calculated with its apparent magnitude and redshift following the formula:
\begin{equation}
    M_i - 5\log h = m_i - 5\log D_{\rm L}(z) - 25,
\end{equation}
where $D_{\rm L}(z)$ is the luminosity distance in units of $h^{-1} \rm Mpc$. We obtain the luminosity of each galaxy $L$ using the formula:
\begin{equation}
    \log (L/h^{-2}\mathrm{L_{\odot}}) = 0.4 \times (4.52 - M_i)\,,
\end{equation}
where 4.52 is the $i$-band absolute magnitude of the Sun.

The galaxy luminosity function measures the comoving number density of galaxies as a function of luminosity, which is one of the most essential tools to characterize the galaxy population. As our galaxy sample covers a large redshift range $0<z<6$, we divide galaxies into 15 redshift bins, each with a bin width $\Delta z=0.4$. We calculate the galaxy luminosity functions at $i$ band in all redshift bins. Note that more comprehensive studies of the luminosity functions based on CLAUDS and HSC data set has been carried out at UV and $U$ band \citep[][Liu et al. in preparation]{Ono2018,Moutard2020,Harikane2021}. Following \citet{Yang2021}, here we use the $V_{\rm max}$ method to calculate luminosity function, 
\begin{equation}
    \phi (M_i) \mathrm{d}\log M_i  = \sum_j 1/ V_{\rm max},
\end{equation}
where the summation is performed for all the galaxies in a given redshift bin.
The $V_{\rm max }$ is the comoving volume within the given redshift bin, which is computed according to the maximum redshift, $z_{\rm max}$, where the apparent magnitude of galaxies can be observed, 
\begin{equation}
    V_{\rm max} = V(\mathrm{min}[z_{\rm max}, z_{\rm bin,up}]) - V(z_{\rm bin,low}),
\end{equation}
where $z_{\rm bin,up}$ and $z_{\rm bin,low}$ are the lower and upper limits of the corresponding redshift bin, respectively.

In Fig.~\ref{fig:lum_gal}, we present the luminosity functions and magnitude-redshift distribution of our sample galaxies selected from different redshift bins at $i$-band. We specifically compare the results for galaxies with and without $K$-correction in calculating the absolute magnitudes of galaxies. Based on the HSC-SSP $grizy$ five band magnitudes and CLAUDS $U$ band magnitude (if available), we calculate the $K$-correction in $i$-band to redshift $z \sim 0$ using the `Kcorrect' model of \citet{Blanton2007}. In the top left panel of Fig.~\ref{fig:lum_gal}, we present the galaxy luminosity functions without the consideration of $K$-correction at seven redshift bins as indicated. The distribution of low redshift galaxies shows a profile consistent with other observations at $i$-band \citep{Blanton2003, Hill2010, Driver2012} as shown with black lines. While at high redshifts, galaxy luminosity functions show some significant enhancements at bright ends.
Shown in the lower left panel of Fig.~\ref{fig:lum_gal} are the luminosity functions which are $K$-corrected to redshift $z=0$. We include a comparison with a relatively high-redshift ($0.7 < z < 0.8$) observation from \citet{Bates2019} shown with black triangles, which matches well with our results after taking $K$-correction into account. There presents a strong evolution in the luminosity functions with $K$-correction at $i$ band taken into account \citep{Mobasher1996}. Overall, the luminosity function with $K$-correction in the lowest redshift bin is similar to that without the consideration of $K$-correction. However in higher redshift bins, the luminosity functions show extraordinary enhanced bumps at the bright ends. The overall behavior deviates from the typical Schechter functional form, and there are galaxies with $M_i<-28$.

The differences can also be clearly seen in the magnitude redshift distributions of galaxies as shown in the right panels of Fig.~\ref{fig:lum_gal}. For the galaxies at $z > 3.5$, there appears a gap for the faint galaxies as the $K$-correction model we used is most applicable to $z \sim 2$ \citep{Blanton2007}. The galaxies at different redshifts have different $K$-correction scatter ranges. 

Note that in this paper, we are using galaxy luminosity as proxy for the halo mass estimation, not trying to provide a coherent and accurate galaxy luminosity function measurement. In addition, as we have separated galaxies into small redshift bins, without $K$-correction only means that we are measuring galaxy luminosity functions in the $observed$ frame around the median redshifts of galaxies in these redshift bins. Furthermore, the derived halo mass is not strongly depend on the luminosity function as we will use an abundance matching method to obtain the halo mass (see more details in next section). 
Given these situations, we decide to use galaxy absolute magnitudes without $K$-correction in this study. 
    
\section{The method and basic quantities} 
\label{sec_method}

In this section we first give a brief description of the method we used to extract groups from our galaxy sample. After the construction of the group catalog, we present some basis quantities of it.   
\subsection{The halo-based group finder}
\label{sec:method}

The group finder we used in this study is a halo-based method developed in \citet{Yang2005,Yang2007}. The galaxy-dark matter connection has been extensively studied in theories (e.g., HOD, stellar to halo mass ratios), which provides the foundation of this group finder. In the first version of the group finder, it was only applicable to galaxies with spectroscopic redshifts, and hence has been applied to the 2dFGRS and SDSS data to search for groups at low redshifts \citep{Yang2005,Yang2007}. Recently, the group finder is improved to be applicable to both spectroscopic and photometric redshifts \citep{Yang2021}. The performance of the new version of the group finder is tested against mock galaxy redshift samples. It turns out that the group finder is stable and reliable to construct group catalogs for galaxy samples with photometric redshift error $\sim 3$\%. This extended version will allow us to probe group contents over large redshift ranges using photometric redshift galaxy data. Here we give a brief description of the main steps in this method \citep[more details can be found in][]{Yang2021}. 

The group finder begins with the assumption that each galaxy is a group candidate. To alleviate the impact of galaxy incompleteness due to the magnitude limit cut, possible impact of galaxy luminosity evolution, and the effect of not taking into account the $K$-correction, here we separate our sample galaxies into 15 redshift bins, each  with an interval of $\Delta z = 0.4$. We measure the total luminosity of each group by summing up the luminosity of all member galaxies, and compute the cumulative group luminosity functions in each 
redshift bin. Meanwhile, the cumulative halo mass function is obtained from the analytic model prediction by \citet{Sheth2001} corresponding to the median redshift of groups in each bin. We determine the mass-to-light ratios of groups in each redshift bin with the cumulative halo mass functions and group luminosity functions using the abundance matching method \citep{Yang2007}. 

Then, each tentative group is assigned a halo mass $M_{L}$ based on the upper mass-to-light ratio using interpolation techniques. With a halo mass, each group can have a halo radius and velocity dispersion along the line-of-sight. The halo radius is defined as 180 times the average matter density of the Universe, expressed as:
\begin{equation}
    r_{180} = 0.781h^{-1}\mathrm{Mpc}\left(\frac{M_{L}}{\Omega_m10^{14}h^{-1}\rm M_{\odot}}\right)^{1/3}(1+z_{\rm group})^{-1},
\end{equation}
where $z_{\rm group}$ is the redshift of the group center.
The line-of-sight velocity dispersion of a dark matter halo is obtained using the fitting function of \citet{van2004} with slight modifications to be suitable to $\Lambda$CDM cosmology with other $\Omega_m$ values:
\begin{equation}
    \sigma_{180}=632 \mathrm{s^{-1} km}\left(\frac{M_L\Omega_m}{10^{14}h^{-1}\rm M_{\odot}}\right)^{0.3224}.
\end{equation}
%The group candidates determine their member galaxies using the halo information. 
The group membership updates begin from the most massive one by taking the luminosity weighted group center as halo center and assuming that the distribution of member galaxies in phase-space follows that of the dark matter particles. The probability of a galaxy to be a member galaxy can be written as:
\begin{equation}
    P_\mathrm{M}(R,\Delta z) = \frac{H_0}{c}\frac{\Sigma(R)}{\bar{\rho}}p(\Delta z),
\end{equation}
where $R$ is the projected distance from the group center, $\Delta z = z - z_{\rm group}$, $c$ is the velocity of light, $\Sigma (R)$ is the projected surface density for a NFW halo \citep{Navarro1997}, $p(\Delta z)$ is a Gaussian function form to describe the redshift distribution of galaxies within the halo \citep[see detail in][]{Yang2021}. Here we have $\sigma= \max (\sigma_{180}, c\sigma_{\rm photo})$ where $\sigma_{\rm photo}$ is the typical photometric redshift error as described by the solid line shown in Fig. \ref{fig:galz}.
Note that in our galaxy sample, if a galaxy has a spectroscopic redshift, we assigned it with a $\sigma_{\rm photo}=0.0001$ value. For this galaxy, the $\sigma=\sigma_{180}$ value will automatically be used, while for the majority galaxies with only
photometric redshifts, $\sigma=c\sigma_{\rm photo}$.

\begin{table*}
    \centering
    \caption{Number, percentage and median $\log$ halo mass of the groups with different number of member galaxies ($N_{\rm g}$) and within different redshift ranges.}
    \label{tab:groupnumber}
    \begin{tabular}{c c c c c c c c }
    \hline \hline
     & {$N_{\rm g} \ge 1$} & {$N_{\rm g} \ge 2$} & {$N_{\rm g} \ge 3$} & {$N_{\rm g} \ge 5$} & {$N_{\rm g} \ge 10$}\\
    redshift & $N_{\rm grp}$ ($\log M_{\rm h})$ & $N_{\rm grp}$ ($\%, \log M_{\rm h})$ & $N_{\rm grp}$ ($\%, \log M_{\rm h})$ & $N_{\rm grp}$ ($\%, \log M_{\rm h})$ & $N_{\rm grp}$ ($\%, \log M_{\rm h})$ \\
    (1) & (2) & (3) & (4) & (5) & (6)  \\
     \hline
     $0 < z < 6$ & 2,232,134 (11.86) & 669,209 (30.0, 12.20) & 402,947 (18.1, 12.32) & 204,099 (9.1, 12.50) & 68,711 (3.1, 12.79)\\
     $z \ge 1$  & 1,617,926 (11.87) & 379,346 (23.4, 12.27) & 178,326 (11.0, 12.49) & 65,034 (4.0, 12.77) & 13,576 (0.8, 13.15)\\
     $z \ge 2$ & 955,893 (11.82) & 132,022 (13.8, 12.22) & 41,815 (4.4, 12.50) & 8,993 (0.9, 12.83) & 919 (0.1, 13.25) \\
     $z \ge 3$ & 374,688 (11.78) & 28,284 (7.5, 12.15) & 5,403 (1.4, 12.40) & 404 (0.1, 12.79) & 5 (0.001, 13.30) \\
     $z \ge 4$ & 59,997 (11.87) & 1,125 (1.9, 12.22) & 89 (0.1, 12.55) & 4 (0.006, 13.00) & 0  \\
     $z \ge 5$ &  6,553 (11.95) & 32 (0.5, 12.20) & 1 (0.01, 12.78) & 0  & 0  \\
    \hline
\end{tabular}
\tablecomments{Column (1) lists the different redshift bins. Column (2), (3), (4), (5) and (6) list the number of groups, $N_{\rm grp}$, with at least 1, 2, 3, 5 and 10 members, respectively. The values of the percentage and log median halo mass in each redshift and richness bin are listed  in the parentheses.}
\end{table*}

\begin{figure}
    \centering
    \includegraphics[width=0.45\textwidth]{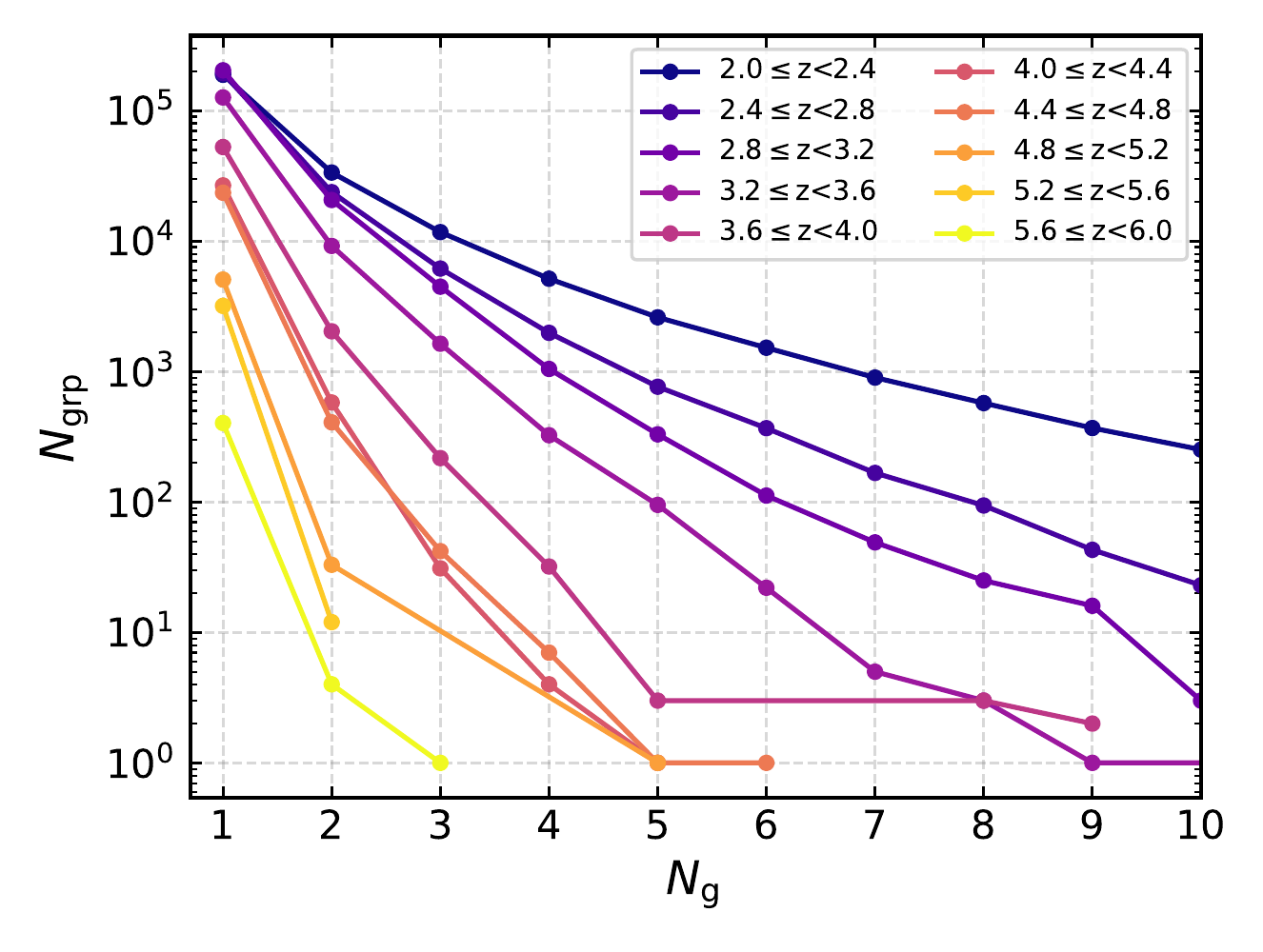}
    \caption{The distribution of the number of groups $N_{\rm grp}$ as a function of the number of member galaxies $N_{\rm g}$. Different colors represent different redshift ranges as shown in the top-right legend. }
    \label{fig:Ng}
\end{figure}

\begin{figure*}
    \centering
    \includegraphics[width=\textwidth]{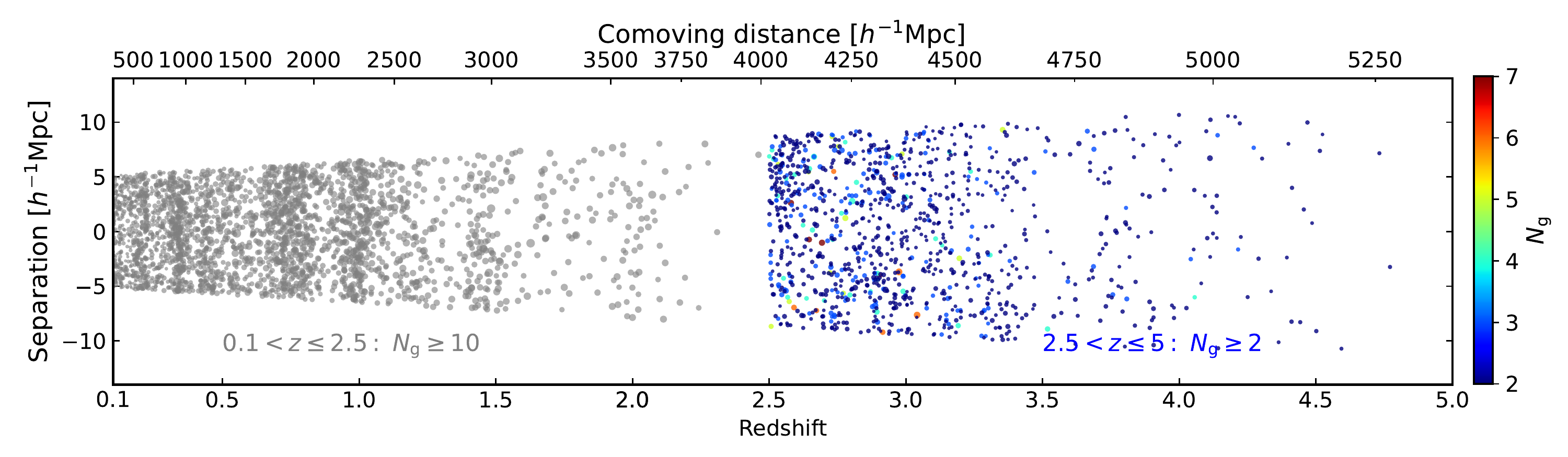}
    \caption{The projected distribution of a selection of groups in a transverse  v.s. line-of-sight direction plane. The groups with at least 10 member galaxies at $0.1 < z \leq 2.5$ are shown with grey dots, while those with at least 2 member galaxies at $2.5 < z \leq 5$ are color coded by the number of their member galaxies. The size of the dot is proportional to the log mass of the group.}
    \label{fig:zMap}
\end{figure*}

Next, we assign galaxies to a candidate group with the judgement between $P_\mathrm{M}(R,\Delta z)$ and $\mathrm{B}\sigma_{180}/\sigma$. If $P_\mathrm{M}(R,\Delta z) \ge \mathrm{B}\sigma_{180}/\sigma$, the galaxy will be assigned to the group. Here, the background value $B$ independent of halo mass perceptively quantifies the threshold of the redshift space density contrast of groups \citep{Yang2005}. We adopt theoretically gauged parameter 10 as the $B$ value during group finding. Decreasing this background value may slightly increase the richness of the groups. The ratio $\sigma_{180}/\sigma$ is used to account for the decrease of density contrast caused by the photometric redshift error. 

After assigning all the galaxies into groups, we update the group centers and luminosities, and recalculate the halo information and find member galaxies again. The iterative stops until there are no more changes for the group memberships. Finally, we start from the beginning to make another iteration, aimed at the convergence of mass-to-light ratios which normally need 3 to 4 iterations.  

\subsection{Survey edge effect} \label{sec:edge}

As we can see in Fig. \ref{fig:dis2D}, the survey geometries of our galaxy catalogs are quite complicated. We follow Y07 to provide a  parameter, $f_{\rm edge}$, to quantify the survey edge effect of the groups. For this purpose, we randomly distribute 200 points within the radius ($r_{180}$) of each halo. Then, we remove those random points that are outside of the survey region according to the mask of the CLAUDS and HSC-SSP data set. For each group, we calculate the number of remaining points, $N_{\rm remain}$, and define the fraction $f_{\rm edge} \equiv N_{\rm remain}/200$ to gauge the volume of the group that lies within the survey edges. About 84 per cent of the groups in our catalog have a $f_{\rm edge}$ value larger than 0.9. For the groups with $M_{\rm h} > 10^{13} \msunh$, $\sim$ 72 per cent of the groups have $f_{\rm edge} > 0.9$. This parameter is provided in our catalog in case some studies need to consider this edge effect.

\subsection{Basic quantities} \label{sec:basic}

After applying the extended halo-based group finder to the selected CLAUDS and HSC-SSP galaxy sample, we obtain a total of 2,232,134 groups. There are 402,947 and 68,711 groups containing at least 3 and 10 member galaxies respectively. As our group catalog covers a wide redshift range, it would be interesting to see those numbers and masses in different redshift ranges. We list the number of groups, the percentage of the groups with respect to the total population, and the median halo mass of groups within different number of member galaxies and redshift ranges in Table.~\ref{tab:groupnumber}. More explicitly, we present the number of groups as a function of the number of member galaxies in different redshift bins in Fig.~\ref{fig:Ng} for groups with redshift $z\ge 2$. With redshift increasing, the number of member galaxies decreases dramatically. At $z \ge 5$, most groups are isolated galaxies indeed, and only 31 groups have two member galaxies, 1 group with three members. 

\begin{figure*}
    \centering
    \includegraphics[width=0.8\textwidth]{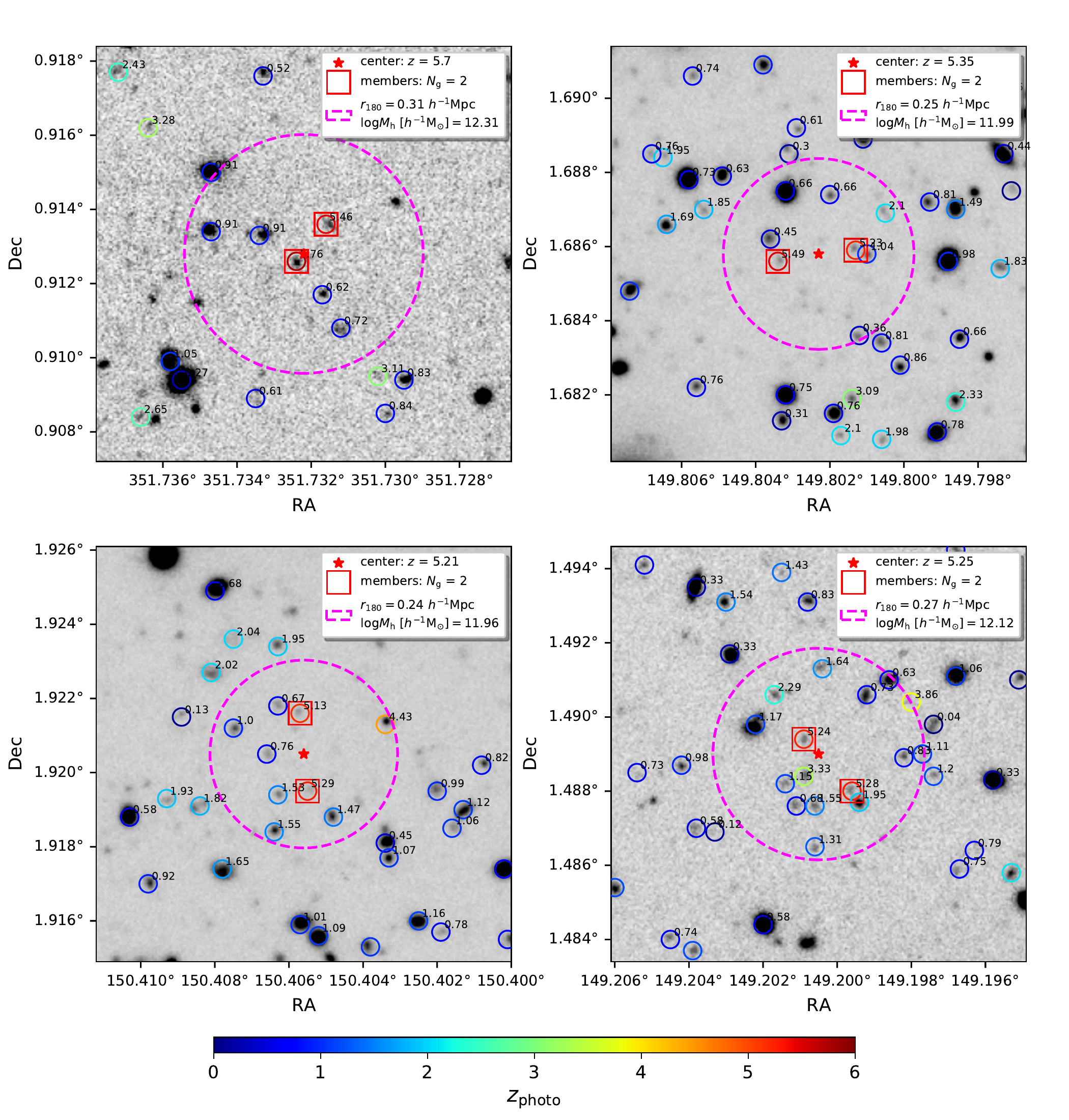}
    \caption{Four cases of groups at $z>5$ in the HSC-SSP PDR2 deep $i$-band images. Galaxies around the group center in our galaxy sample are marked with solid circles color coded by $z_{\rm photo}$ of which values are written on the upper right corners of their positions, while the member galaxies are indicated with red squares. The center and halo radius of the groups are presented with red stars and dashed magenta circles, respectively. Note that the center here is the luminosity weighted center as defined in text. The information of the group including its redshift $z$, number of member galaxies $N_\mathrm{g}$, halo radius $r_{180}$ and mass $M_\mathrm{h}$ are labelled on the top-right corner of each panel.}
    \label{fig:pairs}
\end{figure*}

As an illustration, we show in Fig. \ref{fig:zMap} the projected distributions of the groups selected from a small sky coverage area.
The $y$-axis and $x$-axis represent the transverse distance in the RA direction from the field center and the line-of-sight distance, respectively. It is quite obvious that the richness of groups decreases as the increase of redshift. Most of those groups with at least 10 members are below redshift $z=2.5$.  

We note that in our galaxy sample, the photometric redshifts maybe contaminated by star light, etc., in certain bands. We particularly examine the distribution of groups at $z \ge 4$ in the images from HSC-SSP PDR2. Most groups are well established as clearly shown in the images and present a peak of galaxy numbers around the group center (e.g., Fig.~\ref{fig:fila}). However, we also find that some member galaxies of groups in the images suffer from the contaminations such as scattered light, satellite trail and long wings of bright stars \citep{Aihara2018, Aihara2019}, even though the processing pipeline \citep{Bosch2018} used in HSC-SSP PDR2 has improved a lot. We believe that the groups determined in these cases are not reliable. Although we have excluded some polluted sources by setting selection flags, a few galaxies located in the contaminated area are hard to classify and were left in our sample, occupying a small fraction of the total galaxy sample. About 17\% (2/12)\footnote{The number of groups containing contaminated members divided by the total number of such groups} and 21\% (7/34) of groups with at least two member galaxies at redshift $5.2 < z < 5.6$ and $4.8 < z < 5.2$, respectively, suffer from this problem. This problem is especially severe for the groups with at least three members at redshift $4.4 < z < 4.8$: 71\% (36/51) of the groups are located at the contaminated area (see also the upper-left panel of Fig. \ref{fig:lum_gal} for the enhanced number of very bright galaxies at redshift $z\sim 4.8$, the hint of such a contamination). Thus, it is better to visually inspect the images before studying these individual groups, especially at particular redshift bin $4.4 < z < 4.8$. Besides, at $4.0 < z < 4.4$, only 4 cases in a sample of 36 groups with at least three members suffer this problem. In general, by investigating the redshift distribution of the contaminated galaxies in the images, we find that most galaxies are at $z \gtrsim 4.4$. We suppose that these contaminated galaxies are prone to be assigned high redshifts during their redshift SED fitting. Nevertheless, the current galaxy sample is already among the best deep photometric redshift catalogs we current have, and PFS observation will provide massive spectroscopic redshifts in the near future in these regions. These contaminated groups are excluded in our following analysis and plots.

\begin{figure*}
    \centering
    \includegraphics[width=0.8\textwidth]{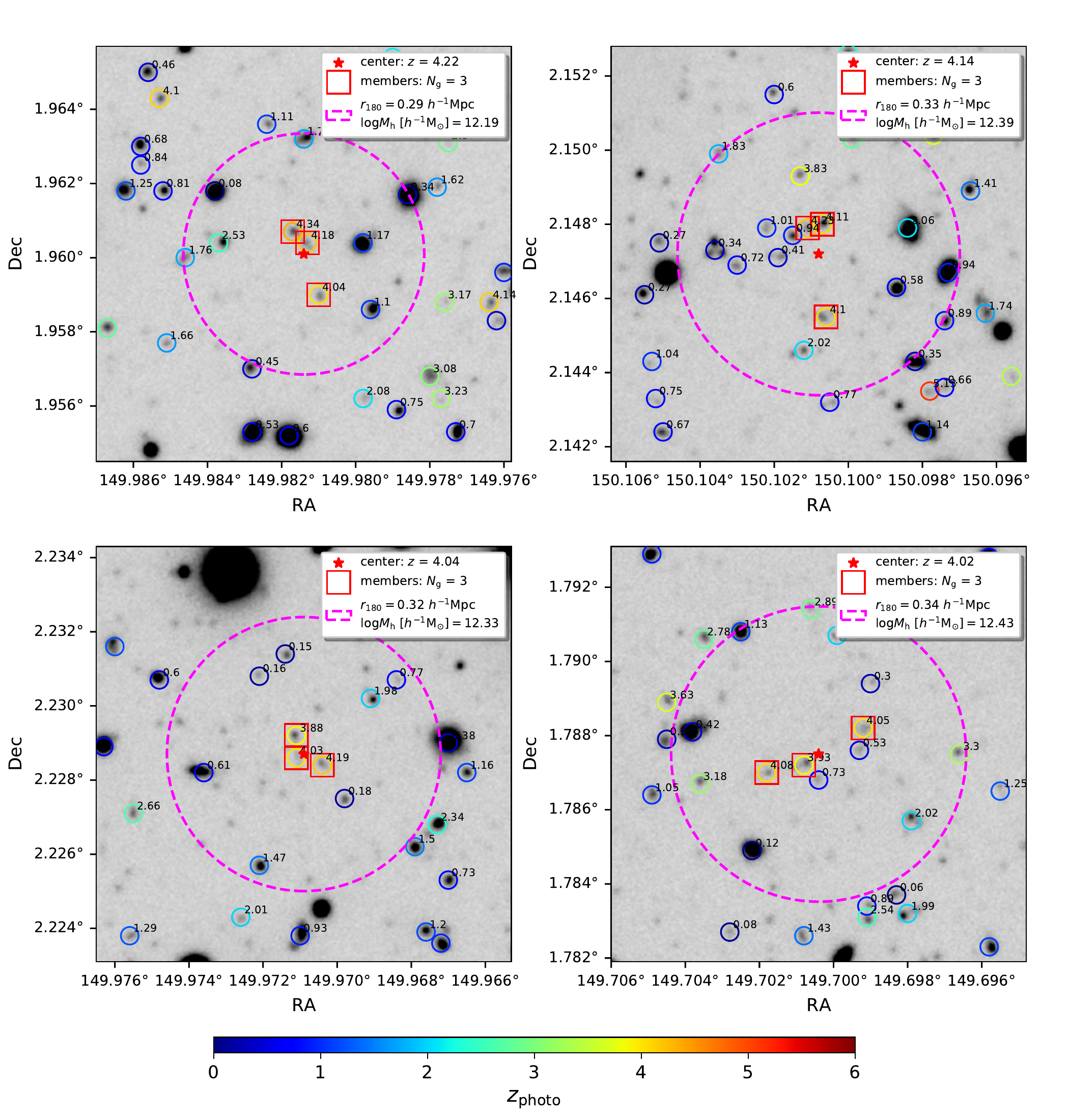}
    \caption{Similar to Fig.~\ref{fig:pairs} but for groups at $z \sim 4$. Here we show groups with 3 member galaxies.}
    \label{fig:triples}
\end{figure*}

\section{Properties of galaxy groups at different redshifts}
\label{sec_group}

We set out to investigate the properties of groups, paying special attention to the evolutionary trend of the groups as a function of redshift. 

\subsection{Galaxy pairs at $z\sim$ 5} \label{sec:pairs}

We start our analysis from high redshift. 
As shown in Fig.~\ref{fig:Ng} and listed in Table \ref{tab:groupnumber}, the vast majority of our groups at $z > 5$ contain only one member galaxy. There are in total only a few tens of galaxy pairs. Comparing to the rich systems at low redshifts, the poor systems found at such high redshift in general have two origins, one is the observation selection effect and the other is the growth of halos. We carry out a rough investigation about the observational selection effect, i.e., faint galaxies may not be observed at high redshifts, on our determination of high redshift group members. We select 100 richest groups at $z \sim 1$ with a median value of 197 members. Then, we move these groups to higher redshifts and estimate the number of remaining group members according to the magnitude limit and without the consideration of the halo evolution. At $z = 5.4$, we find most groups possess about two members, which is roughly consistent with our expectation. However, taking into account the strong evolution of galaxy luminosity functions at redshift $\ga 3.6$, those richest systems should contain more bright galaxies than the ones been observed, indicating a quite
strong growth or accretion of other member galaxies at a later stage. 

We present four of these groups at $z > 5$ in Fig.~\ref{fig:pairs}. The images are taken from the HSC-SSP PDR2 deep observations at the $i$-band. 
We mark galaxies around the group center in our galaxy sample with solid circles color coded by $z_{\rm photo}$ of which values are written on the upper right corners of their positions. We use red squares to indicate the member galaxies. The center and halo size (with radius $r_{180}$) of each group are presented with a red star and dashed magenta circle, respectively. The information of the group including its redshift $z$, number of member galaxies $N_\mathrm{g}$, halo radius $r_{180}$ and mass $M_\mathrm{h}$ are labelled on the top-right corner of each panel.

As the group masses are estimated using the halo abundance matching method \citep[see][for an illustration of the halo mass functions at different redshifts]{Yang2012}, 
the masses of these groups are around $\sim 10^{12}\ h^{-1} \rm M_{\odot}$ with a mean comoving radius of $\sim 0.28\ h^{-1}\rm Mpc$. By checking the neighboring galaxies, we can find that the member galaxies are well determined by our group finder, i.e., no galaxies are seen with similar redshifts near the groups which might be missed. 

Comparing to the halo radius, the member galaxies here seem to have relatively large separations, meaning that at such high redshift, galaxy or halo major mergers are not frequent. 
In addition, the galaxy pairs in most cases do not show significant differences, i.e. distinction between one bright central galaxy (BCG) and the other faint satellite galaxy (FSG). Quite interestingly, evidence has shown that proto-BCGs at $z\sim1.6$ have formed at high-redshift through equal-mass mergers of massive galaxies \citep[see, e.g.,][]{Sawicki2020}. Theoretical models \citep[e.g.,][]{DeLucia2007,Contini2015} also predict such merger formation scenarios. Thus, we speculate that the galaxy pairs at $z \sim\ 5$ might be the progenitors of those equal-mass mergers, of which their later descendants are ultra-massive quiescent galaxies at lower redshifts \citep{Sawicki2020}.

\begin{figure*}
    \centering
    \includegraphics[width=0.8\textwidth]{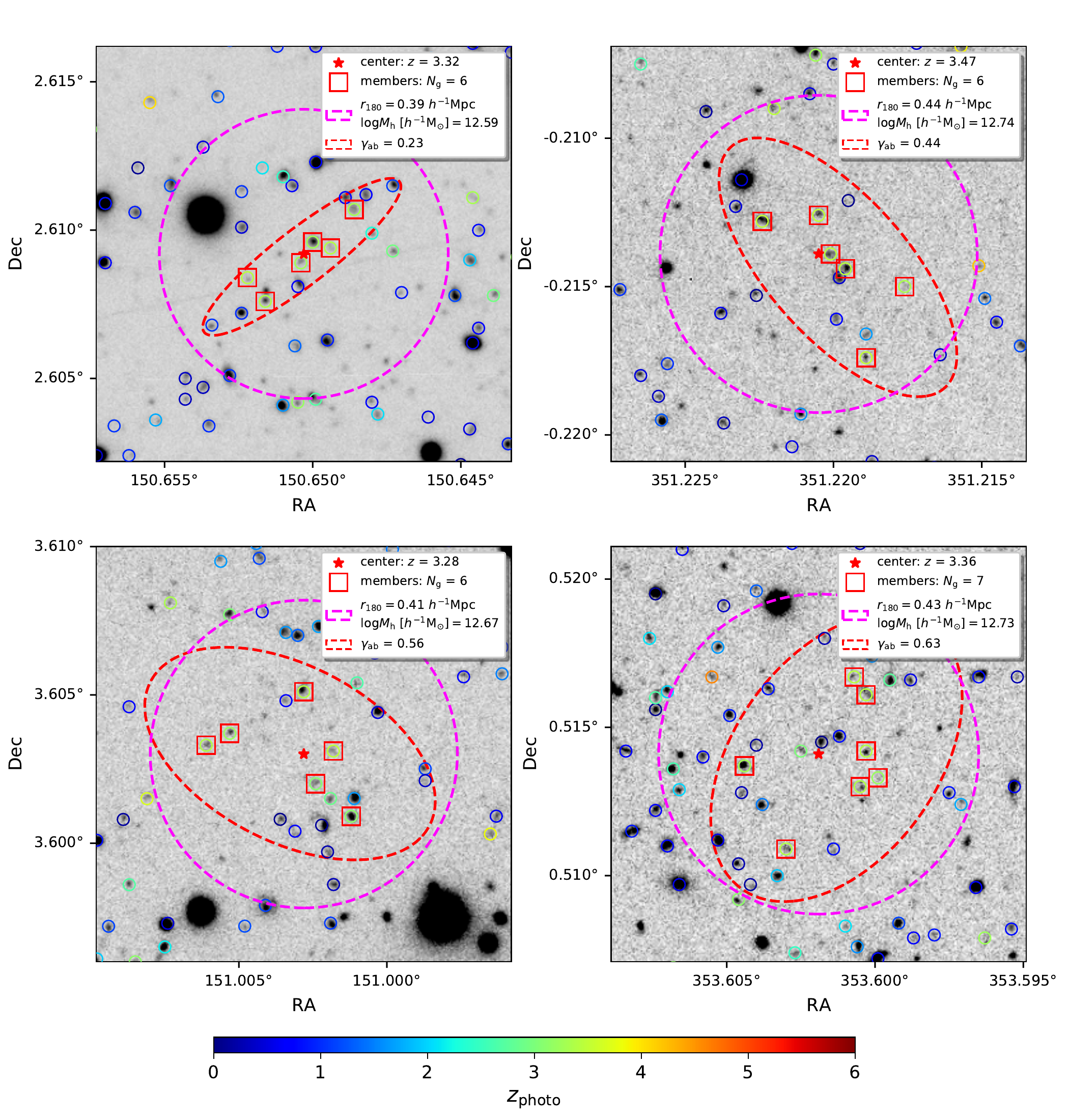}
    \caption{Similar to Fig.~\ref{fig:pairs} but for groups at $z \sim 3$. The red dashed ellipse indicates the 2$\sigma$ coordinate distribution of member galaxies. The ratio between the minor and major axis of red ellipse is written on the right corner of legend. For clarity, we omit the redshift values indicated in previous figures.}
    \label{fig:fila}
\end{figure*}

\subsection{Triple galaxy systems at $z\sim$ 4} \label{sec:triples}

As redshift decreases to $z\sim$ 4, we have more galaxies in our sample but the galaxies determined as members of groups are still small. Apart from four groups with at least 5 members, the most massive groups at this redshift range roughly have 3 members. 
Similar to Fig. \ref{fig:pairs}, we show in Fig.~\ref{fig:triples} the galaxy distributions around four typical groups with 3 members at $z\sim 4$. These groups with halo mass around $\sim 10^{12.3}\ h^{-1} \rm M_{\odot}$ have a mean comoving radius of $\sim 0.32\ h^{-1}\rm Mpc$.

Here again, compared to the member galaxies enclosed in the magenta dashed circles, there are no obvious galaxies at similar redshifts being missed by the group finder. Comparing to those galaxy pairs at redshift $z\sim 5$, here the triple galaxy systems, have the following features: 
\begin{itemize}
    \item Among the galaxies triplets, a pair of galaxies seem to be very close, and the third is somewhat far away. Within the close pairs, there is one galaxy with a relatively higher luminosity (deeper grayscale compared with other galaxies), i.e., the distinction between BCG and FSG starts to be significant. This might be due to the different star formation efficiency associated with the halo potential, gas reservoir and cooling rate. 
 \item Associating these triple systems with those galaxy pairs with similar abundance at higher redshift whose member galaxies are quite similar, this might indicate that the faint galaxy (subhalo) in the close pair might be accreted at quite early in the halo, but not been observed at higher redshift due to the flux limit.
 \item The faint galaxy in the close pairs is about to, but not yet, merged to the brighter galaxy around this redshift. Since it is quite fainter than the primary galaxy, this might correspond to a minor merger.
\end{itemize}

\subsection{Distribution of member galaxies at $z\sim$ 3} \label{sec:orientation}

When redshift decreases to $z\sim 3$, the groups possess more member galaxies. Overall, at $z \ge 3$, there are more than 400 groups with at least 5 member galaxies. There are about 144 groups with at least 6 members.
In Fig.~\ref{fig:fila}, we present four cases of galaxy distributions around groups at $z \sim 3$. Three of them have 6 member galaxies and one has 7 members. The mass of groups is in the range of $10^{12.59}$ to $10^{12.74}$ $h^{-1} \rm M_{\odot}$ with a mean virial radius ($r_{180}$) of $\sim$ 0.42 $h^{-1} \rm Mpc$.

Comparing to the galaxy triples at redshift $z\sim 4$, here the galaxy groups have the following features: 
\begin{itemize}
    \item There are one or two prominent brightest galaxies in each group. They are quite distinct from other galaxies, which might indicate that these galaxies, at least some of them, may had devoured their closest neighboring FSGs and thus became the dominant BCGs.    
  \item In cases there are two prominent brightest galaxies in a group, they usually have quite large separation with or without their associated faint galaxy accompanies.  Such a feature may indicate that it is a relatively newly merged system. 
\end{itemize}

The one or two galaxies prominently brighter than other members in a group can be related to the ``luminosity gap" first described by \citet{Ostriker1975}. As dynamical friction is the strongest for the most massive galaxy in a cluster, the massive galaxies tend to sink into the center of halos more quickly, which means that they merge fairly rapidly before the lower-mass galaxies join them. This results in a galaxy population in the halo that contains one or two very massive galaxy, plus all the lower-mass galaxies. Thus, the most massive galaxy looks like an outlier compared to the rest of the population.

\begin{figure}
    \centering
    \includegraphics[width=0.45\textwidth]{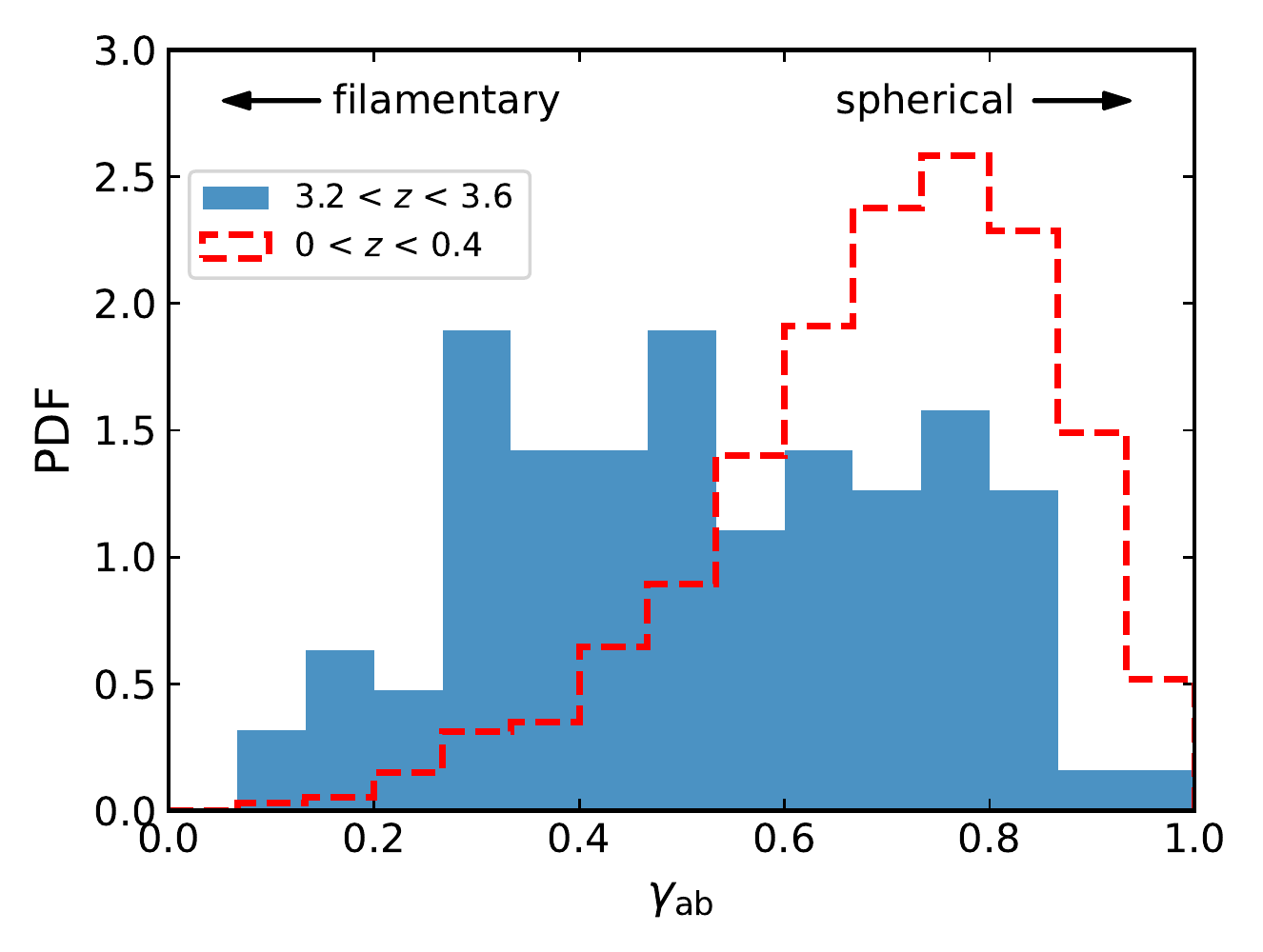}
    \caption{ The probability density function (PDF) of the coordinate ratio, $\gamma_{\rm ab}$.
    The blue filled and red dashed histograms respectively represent the groups at $3.2 < z <3.6$ and at $0 < z < 0.4$ with only five brightest member galaxies. The groups in the low redshift bin are selected according to the halo masses of those groups at $3.2 < z <3.6$. We only consider the brightest 5 member galaxies during the calculations. }
    \label{fig:gamma}
\end{figure}

\begin{figure*}
    \centering
    \includegraphics[width=0.8\textwidth]{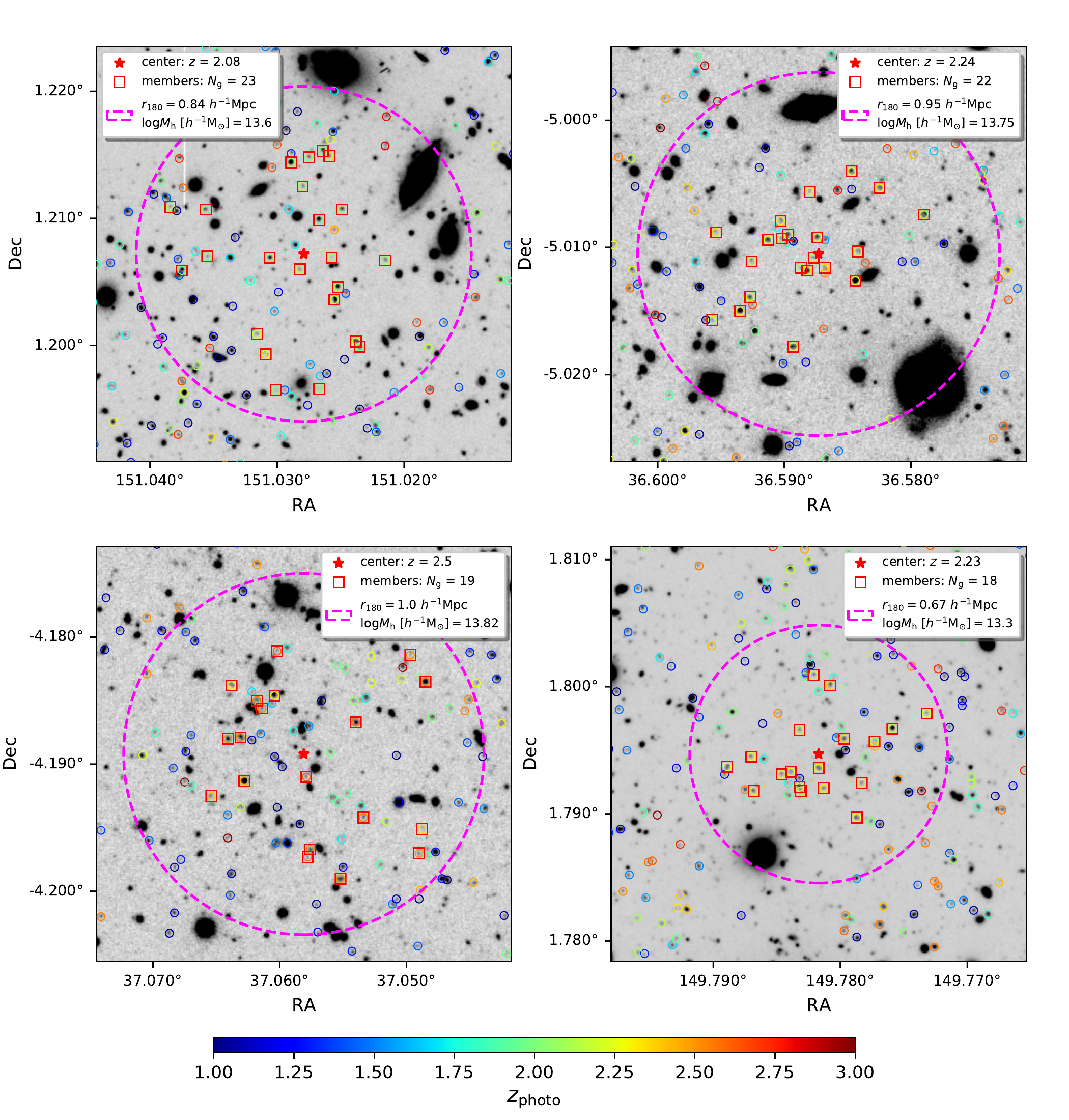}
    \caption{Similar to Fig.~\ref{fig:pairs} but for groups at $z \sim 2$. Note we only show the galaxies at $1 \le z \le3$ marked with the same range color bar.}
    \label{fig:grp_z2}
\end{figure*}

In addition to these case-by-case investigations, we also statistically quantify the distribution of member galaxies in groups since we have a relatively sufficient number of member galaxies. We try to explore the shape of groups indicated by the positions of their member galaxies. Theoretically, the environment of halos is usually divided into four classes: voids, filaments, sheets and clusters (referred as the cosmic web), while filaments are generally considered to form earlier \citep[e.g.,][]{Bond2010,Cautun2014}. At a given redshift, the massive halos are preferentially located in the cluster environments, while low mass halos tend to locate in the filament environments \citep[e.g.][]{Yang2017}. In cases that the shape of the groups are correlated with their surrounding environments, we would expect that the groups with the same halo mass at different redshift would display different shapes. 

Here we directly use the coordinates of member galaxies of groups to quantify their shapes. More explicitly, we take the declination (Dec) and right ascension (RA) relative to the group center as two parameters, while RA is multiplied by $cos$(Dec) to correct the projection effect at high latitudes. Then, we calculate the 2$\sigma$ scatters of these two parameters, ($\Delta \rm Dec$, $\Delta {\rm RA}cos$(Dec)), among the member galaxies with the assumption of 2-dimension Gaussian distribution, which can be shown as an ellipse. We adopt the ratio between the minor and major axis of the scatter ellipse, $\gamma_{\rm ab}$, to indicate the distribution of member galaxies or the shape of groups. In the case of a filamentary distribution, $\gamma_{\rm ab}$ tends to approach zero. In Fig.~\ref{fig:fila}, we show the 2$\sigma$ ellipses with red dashed curves centered in the group center and mark the $\gamma_{\rm ab}$ value on the legend. The four cases with $\gamma_{\rm ab}$ varying from $0.23$ to $0.63$, representing an elongated to relatively spherical shapes of groups  at $z \sim 3$. 

Statistically, we show in Fig.~\ref{fig:gamma} the number of $\gamma_{\rm ab}$ distribution for  groups with $N_{\rm g}=$ 5 members at redshift $3.2 < z < 3.6$, and compare them to a reference group sample at $0 < z < 0.4$ with at least 5 members. Here the reference sample is generated by selecting groups at redshift $0 < z < 0.4$ according to the halo masses in the primary group sample at redshift $3.2 < z < 3.6$. Then for each reference group, we only keep the brightest 5 member galaxies for our consideration. The numbers of groups at $3.2 < z <3.6$ and $0 < z < 0.4$ are 95 and 3983, respectively. In the high redshift sample, $\gamma_\mathrm{ab}$ mainly distributes between 0.3 and 0.8 without a clear peak, while the $\gamma_\mathrm{ab}$ for groups at redshift $0 < z < 0.4$ are concentrated at $\sim$ 0.75. By comparing the distribution of $\gamma_\mathrm{ab}$ in the low and high redshift samples, we infer that the groups at early times are prone to form with a filamentary shape, whereas the groups at late times mainly have a more spherical shape. This is consistent with the formation of LSS under the hierarchical framework that filaments are formed earlier than clusters where galaxies are distributed more spherically than those in the filaments.

% \begin{table*}
%     \centering
%     \caption{The list of protoclusters from other observation data.}
%     \label{tab:procluster}
%     \begin{tabular}{c c c c c c c c }
%     \hline \hline
%      main object & ra & dec & redshift & member & survey & region & paper\\
%      \hline
%      AzTEC-3 & 150.08625 & 2.58933 & 5.3 & 11 &  Keck-II & COSMOS & \citet{Capak2011}\\
%       PHz G237 field & 150.47363 & 2.32747 & 2.16 & 38 & Subaru & COSMOS & \citet{koyama2021}\\
%      0052 & 13.62417 & -23.85864 & 2.86 & 57 & $3-4 \times 10^{14} M_{\odot}$ &  & \citet{Venemans2007}\\
%      0943 & & & 2.92 & 65 & $3-4 \times 10^{14} M_{\odot}$ &  & \citet{Venemans2007}\\
%      1338 & & & 4.11 & 33 & $6-9 \times 10^{14} M_{\odot}$ & &  \citet{Venemans2007}\\
%      \hline
%      \end{tabular}
% \end{table*}

\subsection{Rich groups at $z\sim$ 2} \label{sec:rich}

As redshift decreases to $z\sim 2$, groups possess more member galaxies as the increasing cosmic star formation rate \citep{Katsianis2021} and merger rate \citep{Ventou2017} at this period. There are 914 groups at redshift $z \sim 2$ with at least 10 members. These groups have a median halo mass of $1.8 \times 10^{13}\ h^{-1}\rm M_{\odot}$. The radius of groups ($r_{180}$) can extend to $\sim 1 \mpch$.

In Fig.~\ref{fig:grp_z2}, we show four rich groups at $z \sim 2$ with member galaxies $N_{\rm g} \sim 20$. Overall, the member galaxies distribute more isotropically compared with the groups at $z \sim 3$ (Fig.~\ref{fig:triples}). In addition, quite different from those groups at redshift $z \ga 3$ where there are almost no galaxies at similar redshift close to their halo boundaries (radius), here we start to see some galaxies with similar redshifts approaching to the host groups. These coeval galaxies frequently appearing around the boundary of groups, will significantly contribute to the growth of groups, i.e. to form clusters, in a later stage.

The evolution of major mergers rate are popularly studied in simulations and observations despite still in debates at  high redshifts. The rate of major mergers investigated in some simulations can be increasingly extended to $z \gtrsim 3$ \citep[e.g.,][]{Stewart2008, Lagos2018}, whereas a few simulations show a steady profile at $z \sim 2-3$ \citep[e.g.,][]{Kaviraj2015, Qu2017,Snyder2017}, which is consistent with the results investigated with photometric and flux-ratio-selected galaxy pairs in observations \citep[e.g.,][]{Bluck2009,Man2012,Duncan2019,Ventou2019}. As we have obtained a rather uniform flux-limited galaxy group sample at high redshift, we can use the separation between member galaxies, as well as the stellar mass growth of member galaxies to probe the merger rates of galaxies in observation, and compare with theortical model prediction \citep[e.g.,][]{Jiang2008, Oleary2021}.
%Our results agree with a picture that the merger rate is not necessarily increases with redshifts \citep[e.g.,][]{Duncan2019, Oleary2021}.

\subsection{Groups/clusters at lower redshifts}

At lower redshifts, there are a total of 68711 groups  with at least 10 members in our group catalog. Here we do not provide further visual inspections of these individual groups/clusters, but try to evaluate our low redshift groups by comparing them with other group/cluster data sets.
Fruitful number of groups/clusters at low redshifts have been found with different methods. Here we compare our data with two sets of them.

\begin{figure}
    \centering
    \includegraphics[width=0.45\textwidth]{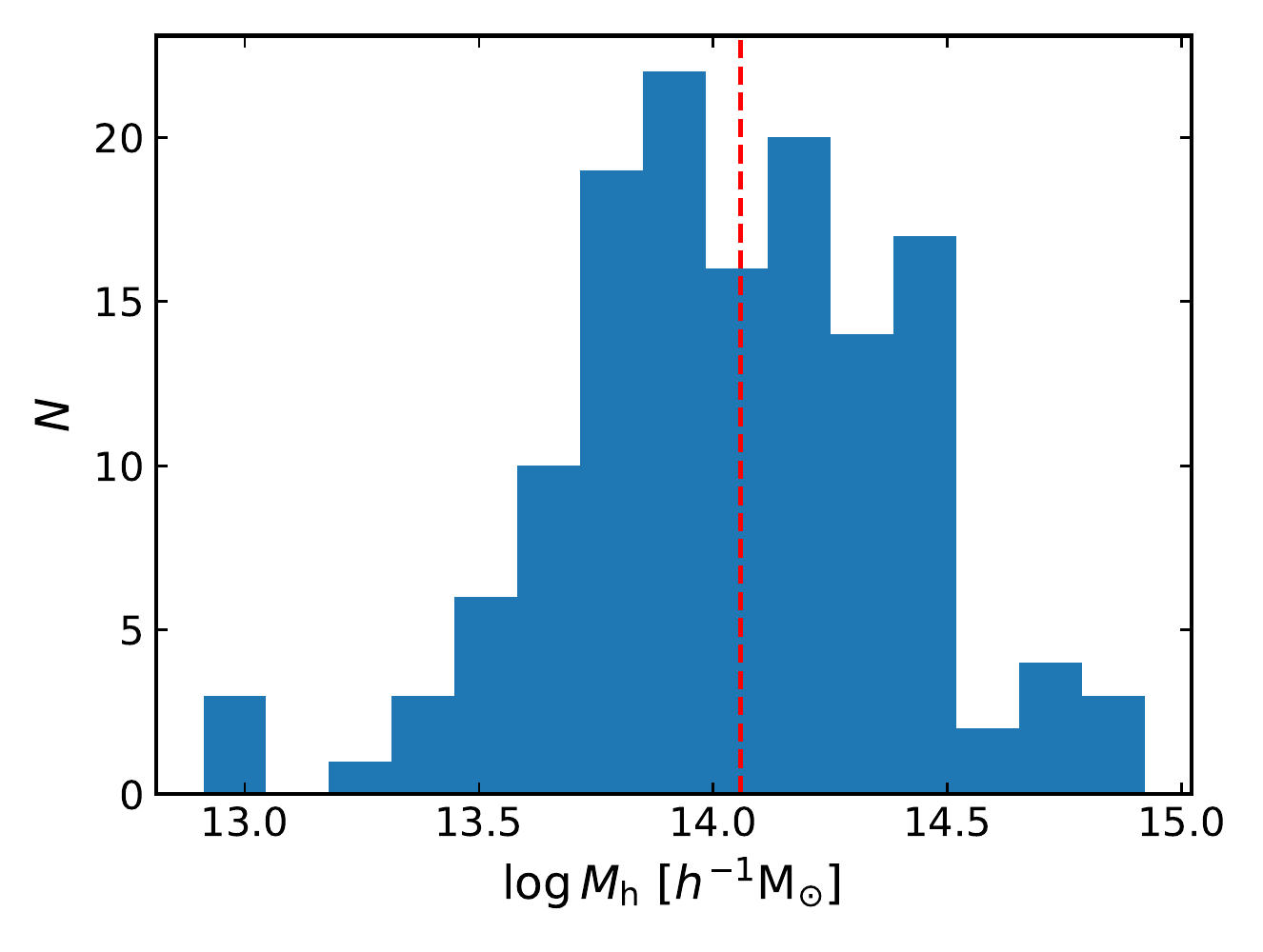}
    \caption{Number distribution for the halo mass of our groups matched to \citet{Oguri2017}. The red line marks the median halo mass, $\sim 10^{14.08} \msunh$.}
    \label{fig:Com_Oguri}
\end{figure}

We use our galaxy groups to match the groups determined in \citet{Oguri2017}, who applied the CAMIRA (Cluster-finding Algorithm based on Multi-band Identification of Red-sequence
galaxies) algorithm \citep{Oguri2014} to the HSC Wide S16A data set. Here we consider their updated data version which used the HSC-SSP PDR2 with photometric redshifts. There are 197 groups at $0.1 < z < 1.2$ in total located in the same fields as ours. As the number density of their groups is roughly above the theoretical curve of $10^{14}\ h^{-1} \rm M_{\odot}$ halos according to the halo mass function \citep{Oguri2017}, they are mostly clusters with mass larger than $10^{14}h^{-1} \rm M_{\odot}$. Before we proceed to match these clusters with our group/cluster sample, we check if the BCGs (also determined as centers) in their cluster sample exist in our galaxy sample. Requiring the photo-$z$ difference to be less than 0.2, we have a total number of 140 clusters remained for our cross check. Afterwards, we search in our group catalog around each of their clusters within a projected 2 $\mpch$ comoving radius and a redshift difference $\Delta z < 0.2$. We take the richest group that fulfils these criteria as its matched counterpart. In Fig.~\ref{fig:Com_Oguri}, we show the number distribution of halo masses in our matched sample. The matched groups are massive with a median of $10^{14.08}\msunh$, which is close to the induced mass of groups in \citet{Oguri2017}. 
The lowest group mass in our matched sample is about $10^{13.0}\msunh$, which is still quite massive. Note here we are using different photometric redshift sources and group detection technique from the ones used in \citet{Oguri2017}, which is the main cause of the differences.

%Then, we match another group sample from \citet{Yang2021}, who constructed a large number of groups with a total $\sim$18,000 $\rm deg^2$ sky coverage of DESI DR8 data set based on the halo-based group finder.  The galaxy sample they used are selected with magnitude $z \le 21$ and redshift $0 < z \le 1$. We use their latest version where they updated the group catalogs with DESI DR9 data set and incorporated more spectroscopic redshifts. \adr{There are $\sim$ 160 thousand groups with $M_{\rm h} > 10^{14} \msunh$. Before matching, we limit these groups by requiring their BCGs exist in our sample and to have photoz differences to be less than 0.2. This filters out 196 groups altogether. } The matching criteria are the same as the match to the sample in \citet{Oguri2017}. \adr{Since the group masses are provided in both \citet{Yang2021} and our study, we show in Fig.~\ref{fig:Com_DESI} the comparison of the halo masses for the matched pairs. }

%\begin{figure}
%    \centering
%    \includegraphics[width=0.45\textwidth]{Com_DESI.pdf}
%    \caption{Number distribution for the halo mass of groups in \citet{Yang2021} matched to our groups with $z < 1$ and $M_{\rm h} > 10^{14} %\msunh$. The red line marks the median halo mass, $\sim 10^{13.73} \msunh$.}
%    \label{fig:Com_DESI}
%\end{figure}

\section{Protocluster candidates}
\label{sec_proto}

As we outlined in Table \ref{tab:groupnumber}, there are 914 groups at redshift $2 \leq z < 3$ with at least 10 members. These groups have a median halo mass of $1.8 \times 10^{13}\ h^{-1}\rm M_{\odot}$. At redshift $3 \leq z < 4$, there are 400 groups with at least 5 members and a median halo mass of $6.1 \times 10^{12}\ h^{-1}\rm M_{\odot}$. At redshift $z\ge 4$, there are 89 groups with at least 3 members and a median halo mass of $3.5 \times 10^{12}\ h^{-1}\rm M_{\odot}$. According to the abundance of these groups and the theoretical median mass growth history of dark matter halos \citep[e.g.][]{Zhao2009, Yang2012}, these groups should {\it on average} be able to grow into clusters with mass $\ga 10^{14}\ h^{-1}\rm M_{\odot}$ at redshift $z=0$. 
However, individually, they might not all grow into clusters \citep[see the discussion in][]{Cui2020}. In this section, we set out to assess the probability of growing into clusters for these group systems.  We refer to those with high possibilities as protocluster candidates. 

For convenience, we separate the groups used to search protocluster candidates into three samples according to their redshifts and number of members: (1) $2 \leq z < 3$ and $N_{\rm g} \geq 10$; (2) $3 \leq z < 4$ and $N_{\rm g} \geq 5$; (3) $ z \geq 4$ and $N_{\rm g} \geq 3$, which are referred to as samples S1, S2 and S3 respectively. We exclude groups which suffer significantly from the survey edge effects including the bright star mask or other contaminations (see Section.~\ref{sec:basic}). Finally, we obtain 761, 343 and 43 groups in S1, S2 and S3 samples. 

% We do not exclude the protocluster candidates with small masks, as we suppose these small masks are randomly and averagely distributed in different directions on the whole sample.

\subsection{Assessment indicators}

In general, galaxy clusters in the local universe are embedded in a virialized dark matter halo with a mass greater than $10^{14} \msunh$. Estimating the predicated halo mass at $z=0$ of a high redshift overdensity is a common way to judge whether the discovered structure is a protocluster or not \citep[e.g.,][]{Chiang2013,Cheema2020,Polletta2021}. However, the methods used to estimate the $z=0$ halo mass usually rely on the overdensity and accurate redshift (volume) measurements \citep[e.g.,][]{Steidel1998, Cucciati2014}, or on clustering measurements \citep[e.g.,][]{Cheema2020}. For our sample, groups are determined with photometric redshifts and their neighboring overdensity may be significantly affected by the photo-$z$ quality. Thus, rather than the overdensity, here we introduce a new set of indicators based on the distribution of neighboring galaxies and groups to quantify their possibilities of being protocluster candidates.

The galaxies we linked to each group are supposed to be mostly located within the halo virial radius ($r_{180}$). Those systems in our S1, S2 and S3 samples can be regarded as the cores of protoclusters as also shown in \citet{Ando2022}, then we check the available matter (or galaxies) surrounding them. If the mass and/or the number of galaxies within and surrounding the group are sufficient to build a redshift $z=0$ cluster with mass $\ga 10^{14}\ h^{-1}\rm M_{\odot}$, we regard it as a protocluster candidate.  
As the forming area of a protocluster is much larger than its already formed group (halo) region at high redshift, we search the galaxy and group distributions around each of the candidate groups within a projected radius and a redshift difference $\Delta z \le 0.1$, i.e., neighboring criteria. The chosen radius criteria for S1, S2 and S3 are $\sim 5$, 6 and $7\mpch$, respectively, which roughly correspond to the effective radius of ``Virgo" type protoclusters at different redshifts defined in \citet{Chiang2013} (see also Fig.3 in their paper). The chosen redshift boundary roughly corresponds to a 3-$\sigma$ scatter of photometric redshift as shown in Fig.~\ref{fig:redz}. 
For convenience, we call the galaxies and groups around the candidate groups fulfil the neighboring criteria (within chosen radius and redshift) as neighboring galaxies and groups in the following analysis.
% according to the effective radius of protoclusters at different redshifts discussed in \citet{Chiang2013}. Although the radius for individual protocluster candidates is diverse, our adopted radius roughly corresponds to a median of radius range for different. 

\begin{figure}
    \centering
    \includegraphics[width=0.45\textwidth]{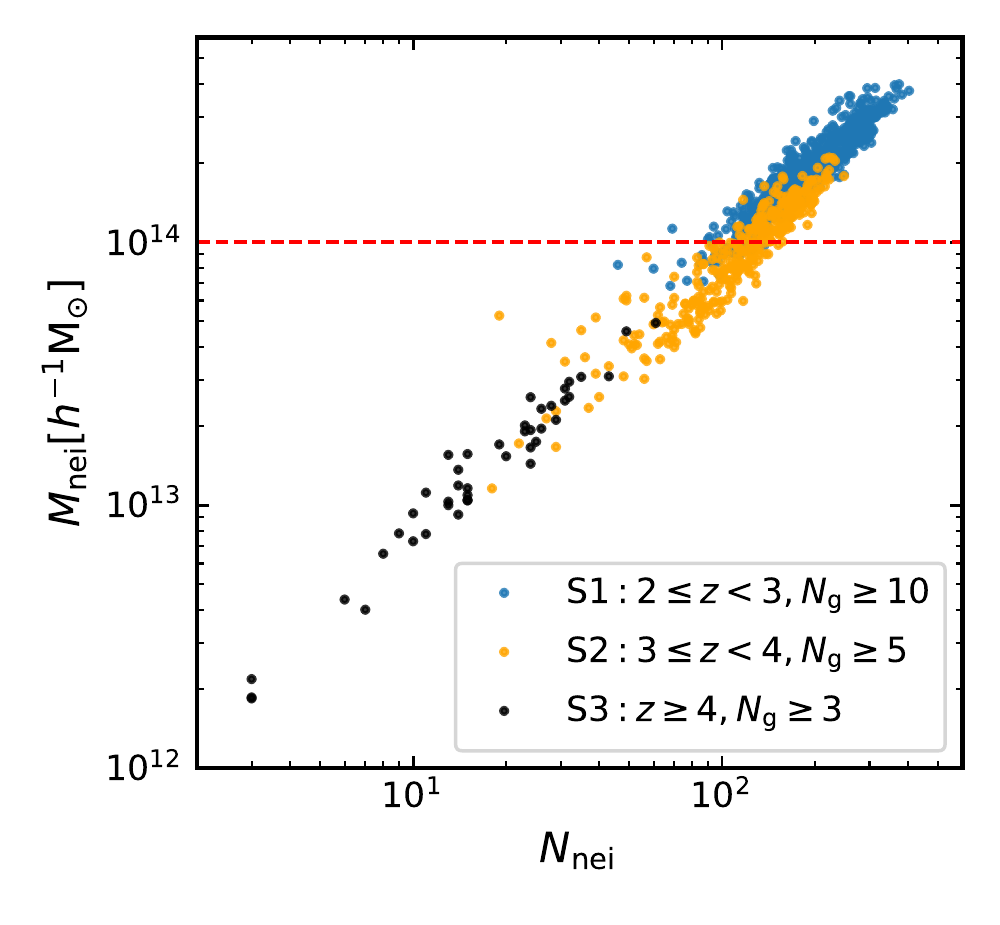}
    \caption{The total halo mass of the neighboring groups verses the total number of neighboring galaxies. The neighboring galaxies or groups are searched within the projected 5, 6 and 7 $\mpch$ comoving radius for the S1, S2 and S3 samples respectively. The groups in these three samples are shown with blue, orange and black dots.  The red dashed line marks the criterion ($10^{14} \msunh$) in $M_{\rm nei}$ for a group with neighboring galaxies considered as a protocluster candidate.}
    \label{fig:Mnei_Nnei}
\end{figure}

\begin{figure}
    \centering
    \includegraphics[width=0.45\textwidth]{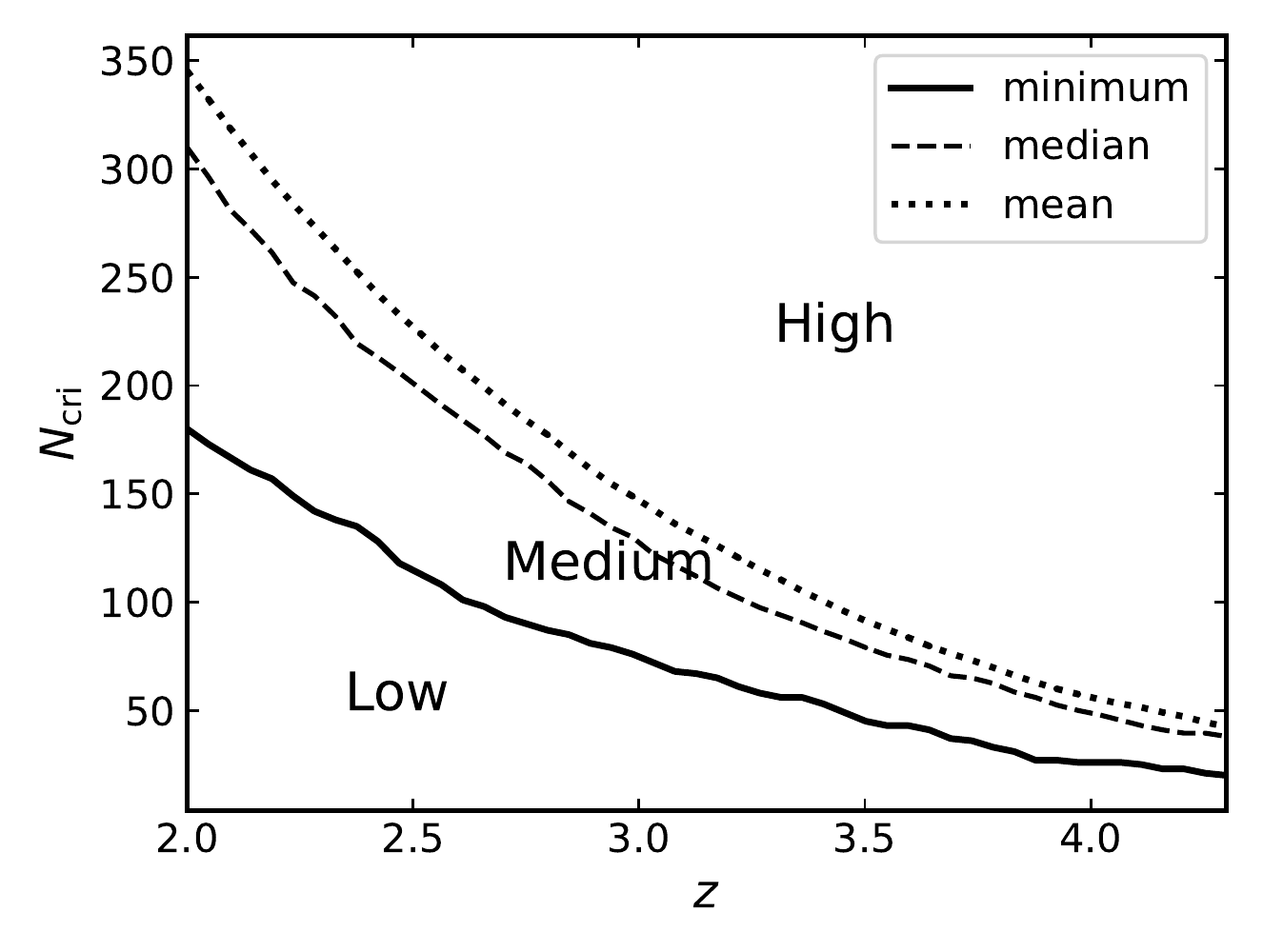}
    \caption{Redshift evolution of the criteria number of neighboring galaxies, $N_{\rm cri}$. $N_{\rm cri}$ is calculated based on the low-redshift massive clusters. The solid, dashed and dotted lines respectively represent the minimum, median and mean criteria number of neighboring galaxies. The protocluster candidates below the minimum $N_{\rm cri}$, between the minimum $N_{\rm cri}$ and mean $N_{\rm cri}$, above the mean $N_{\rm cri}$ are respectively regarded having a low, medium and high probability to grow into clusters respectively.}
    \label{fig:Ncri}
\end{figure}

\begin{figure*}
    \centering
    \includegraphics[width=\textwidth]{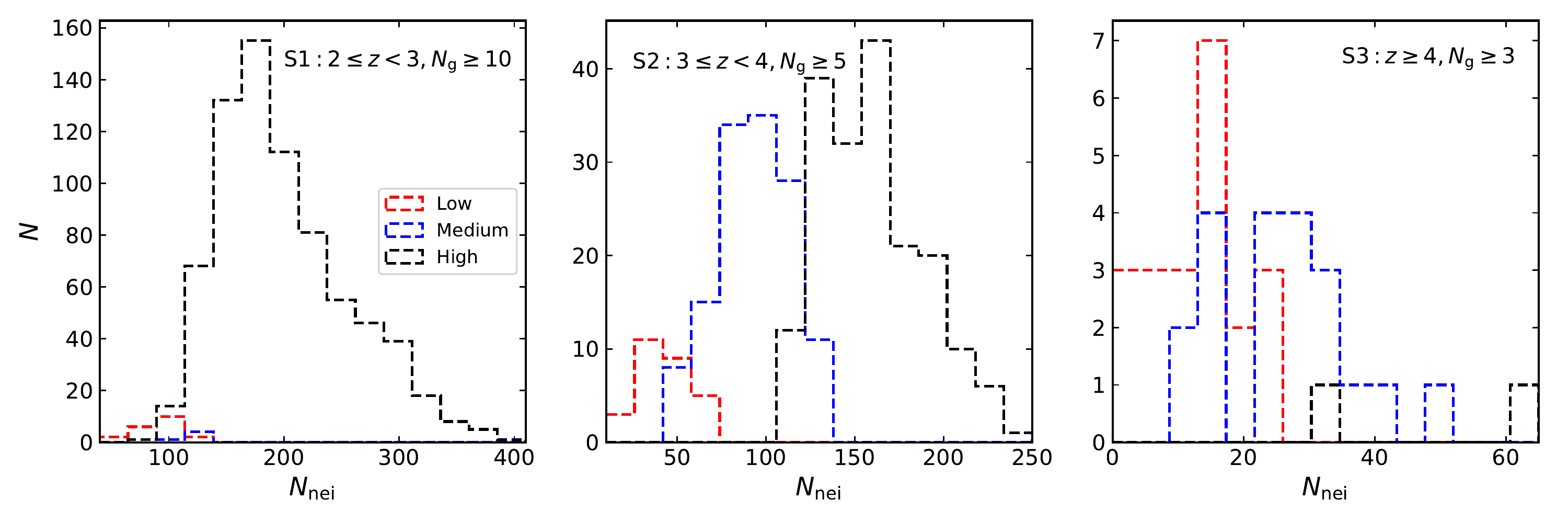}
    \caption{The number distribution of protocluster candidates as a function of the number of neighboring galaxies, $N_{\rm nei}$. The left, middle and right panel show the protocluster candidates in sample S1, S2 and S3 respectively. The red, blue and black dashed lines respectively represent the protocluster candidates with a low, medium and high probability to grow into clusters.}
    \label{fig:Nnei}
\end{figure*}

The first indicator we set out to use is the total halo mass of the neighboring groups, $M_{\rm nei}$. Because of the photometric redshift error as well as the velocity dispersion of member galaxies, some neighboring groups only contribute a fraction of their member galaxies according to our neighboring criteria, we calculate $M_{\rm nei}$ by taking into account the luminosity fraction of the satisfied member galaxies in their groups. 
%Both $N_{\rm nei}$ and $M_{\rm nei}$ are computed within a projected comoving radius, which corresponds to our chosen radius criterion for S1, S2 and S3. As these two parameters rely on the distribution of neighboring galaxies, we throw the groups surrounded with large masks or located near the edge of fields. 
If thus calculated $M_{\rm nei}$ is larger than $10^{14} \msunh$, we can in general directly regard this group as a protocluster candidate. 
However, since we are using a flux-limited galaxy sample, we can only detect groups whose member galaxies can pass the limit, which results in a halo mass limit in calculating the $M_{\rm nei}$. It is not straightforward to use $M_{\rm nei}$ as the criteria to assess the probability of a group being a protocluster candidate, especially at very high redshifts.

The next indicator we use is the total number of neighboring galaxies, $N_{\rm nei}$. This parameter overall reflects the available neighboring galaxies that can be accreted to the target group in supporting its growth toward a cluster at a later time. Note that here we do not intend to provide the probability of being  protocluster candidates for all the groups,  but rather to provide a number of neighboring protocluster candidates in our catalog based on a small selection of richest groups at different redshift bins. Since the halo masses are estimated from the ranking of total group luminosity, there should be some intrinsic correlation between $M_{\rm nei}$ and $N_{\rm nei}$. As an illustration, we show in Fig.~\ref{fig:Mnei_Nnei} the distributions of our rich group systems in a $M_{\rm nei}$ v.s. $N_{\rm nei}$ plane. They actually show a strong and continuous linear correlation across different redshift ranges, which indicates that our defined quantities $N_{\rm nei}$ can be regarded as a substitution of halo mass estimation. Most groups in S1 can directly be determined as protocluster candidates according to the value of their $M_{\rm nei}$, whereas none of clusters can be regarded as protocluster candidates only judged by $M_{\rm nei}$ in S3. Nevertheless, even if $M_{\rm nei}$ is smaller than $10^{14} \msunh$, it does not necessarily mean it can't form a cluster at a later time, since our calculation of $M_{\rm nei}$ suffers from the halo mass incompleteness and these high-redshift halos are still forming with the process of accumulating its mass. In the next subsection, we will provide the details of using 
$N_{\rm nei}$ as the assessment indicator by properly taking into account the survey magnitude limit and halo evolution effects.

\begin{figure*}
    \centering
    \includegraphics[width=\textwidth]{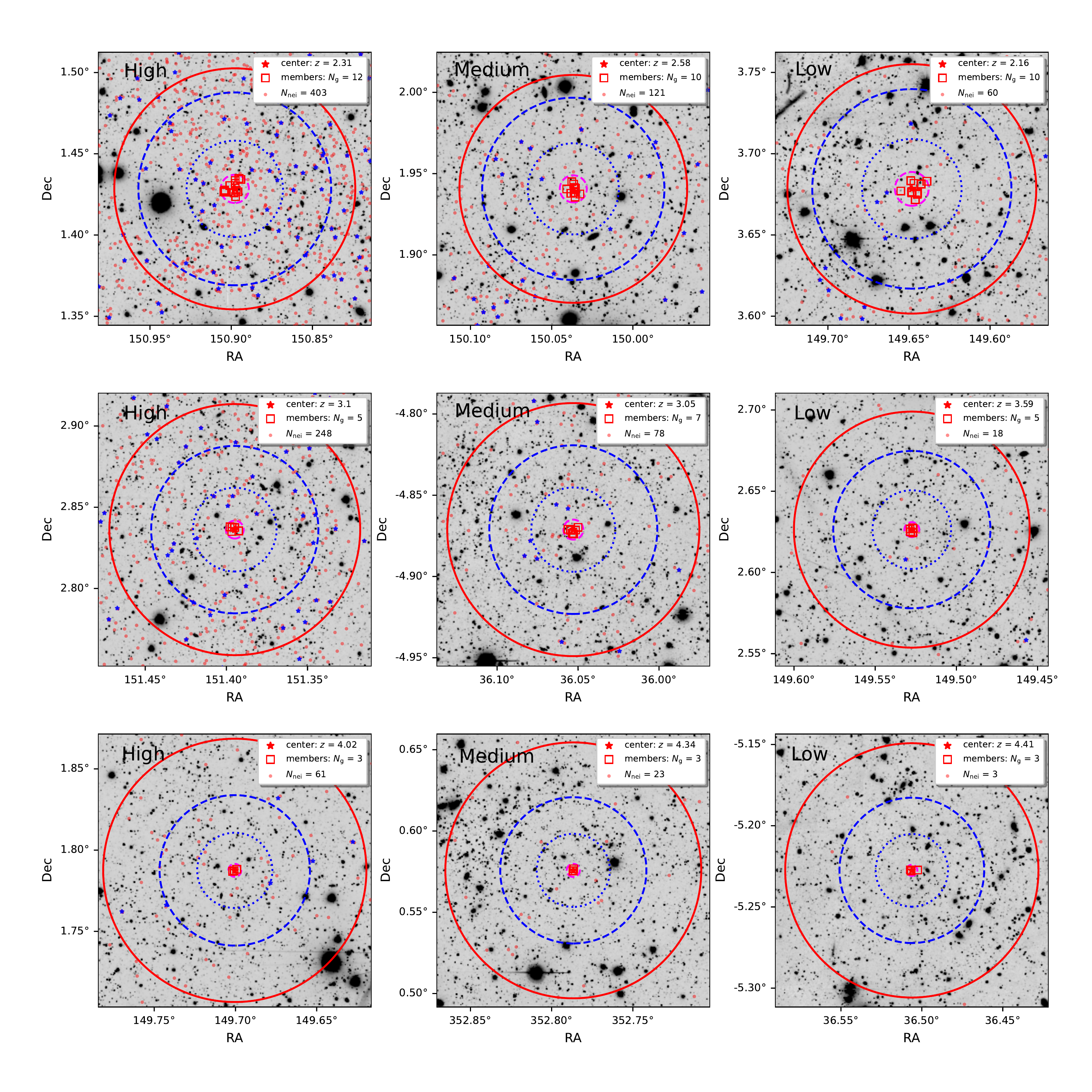}
    \caption{Cases for the distribution of neighboring galaxies and groups in the HSC-SSP PDR2 deep $i$-band images. The left, central and right column panels shows the protocluster candidates under a high, medium and low probability to grow into clusters respectively. The top, middle and bottom row panels indicate the cases in the sample S1, S2 and S3. The neighboring galaxies are marked with red dots, while members are presented with red squares. The positions of neighboring groups with redshift differences to the central group less than 0.1 and having at least two member galaxies are indicated with blue stars. The range of central group ($r_{180}$) is presented with dashed magenta line. The chosen radius to count $N_{\rm nei}$ for the protocluster candidates in the top, middle and bottom panels are respectively 5, 6 and 7 $\mpch$ comoving distance shown with red lines. A radius of 2 and 4 $\mpch$ comoving distance circles are presented as the dotted and dashed blue lines respectively. The value of $N_{\rm nei}$ is labeled in the right-top legend.}
    \label{fig:eg_Pc}
\end{figure*}

\subsection{Finding the protocluster candidates}

In this section, we judge if the target group is a protocluster or not according to the criteria $N_{\rm cri}(z)$ that can be obtained in the following way. We first select clusters at redshift $z \le 0.3$ with mass $\ga 10^{14} \msunh$. We obtain 36 clusters after this selection. Next, we select neighboring galaxies around each of these clusters within a projected maximal distance of their member galaxies and a redshift difference $\Delta z \le 0.1$. We then move these clusters and their neighboring galaxies to higher redshift. According to the magnitude limit, we applied to our galaxy sample, we discard member galaxies that can not make the survey magnitude limit. We can thus obtain $N_{\rm cri}(z)$ by counting the remaining galaxies. When moving low-redshift clusters to high redshifts, we do not take the galaxy evolution into account as the luminosity functions (top left panel in Fig.~\ref{fig:lum_gal}) shows a weak evolution at $z \lesssim 3$. We calculate $N_{\rm cri}$ from each selected cluster and show the minimum, median and mean $N_{\rm cri}$ as a function of redshift in Fig.~\ref{fig:Ncri}. In general, $N_{\rm cri}$ decreases with the increasing of redshift. A cluster is expected to possess at least 180 neighboring galaxies at $z \sim 2$ and 26 at $z \sim 4$. We quantify the probability of a protocluster candidate to grow into a cluster based on the $N_{\rm nei}$ value relative to criteria $N_{\rm cri}$ at the corresponding redshifts. We classify this probability of growth into three levels. 
Specifically, a protocluster candidate with $N_{\rm nei}(z) < N_{\rm cri, min}(z)$ is considered to have a \emph{low} probability to grow into a cluster, and $N_{\rm cri, min}(z) \le N_{\rm nei}(z) < N_{\rm cri, mean}(z)$ and $N_{\rm nei}(z) \ge N_{\rm cri, mean}(z)$ for a $\it{medium}$ and \emph{high} probability respectively. Besides, we consider the protocluster candidates under a high probability as long as its $M_{\rm nei}$ is larger than $10^{14} \msunh$.  

The number distribution of protocluster candidates in the S1, S2 and S3 samples with the division of three probabilities is shown in Fig.~\ref{fig:Nnei}. Overall, the protocluster candidates with a large $N_{\rm nei}$ have higher probability to grow into a cluster. For the S1 sample, most candidates have a high probability, within which quite a large number of these candidates already possess $M_{\rm nei}$ larger than $10^{14} \msunh$. For the S2 sample, the number of protocluster candidates with a medium probability increases compared with those in the S1 sample. At $z \geq 4$, only a less protocluster candidates are under a high probability. The overlapped area among different probabilities is a result of different redshifts of groups judged by $N_{\rm cri}(z)$. 

As an illustration, we show each cases for the protocluster candidates that have a high, medium and low probability of evolving into a cluster in the left, middle and right column panels of Fig.~\ref{fig:eg_Pc}. The top, central and bottom row panels correspond to the cases at different redshift range selected from sample S1, S2 and S3 respectively. The neighboring galaxies are marked with red dots. The radius adopted to count $N_{\rm nei}$ is shown with red circle. Overall, the protocluster candidates with a high probability possess sufficient and dense distribution of neighboring galaxies, whereas the systems with a low probability present a sparse galaxies distribution. With the increasing of redshift, the total number of neighboring galaxies, $N_{\rm nei}$, decreases as expected. In addition, the protocluster candidates in a dense environment seem to have more group member galaxies. 
In summary, the groups in our catalog have a large probability to grow into clusters, especially for the groups at $2 \leq z < 3$ with at least 10 member galaxies. In the group catalog, we have also provided the related $N_{\rm nei}$ values for those candidate groups in consideration.

% \citep{Lemaux2018} searched proto structures within $r_{\rm proj} \le 2$ Mpc for the sample at $1.5\ge z \ge 5$, and a proto-strucutre at $z \sim 4.57$ spanning 7.5 Mpc.% Thus there are few contemporaneous galaxies surrounding the center of protoclusters. Moreover, the edges ofprotoclusters are not clear even beyond 4 $h^{-1}$ Mpc.  

\section{Summary} \label{sec_summary}

We construct a galaxy group and protocluster candidates catalog by applying the extended halo-based method of \citet{Yang2021} to the CLAUDS and HSC-SSP joint deep data set. The extended version group finder allows dealing with a large number of galaxies with both photometric and spectroscopic redshift. In total, we obtain 2,232,134 groups at $0 < z < 6$, of which 41,815 groups have at least 3 member galaxies. We specifically explore the properties of groups at different redshifts by showing the distributions of galaxies in the $i$-band images from HSC-SSP PDR2 deep observation. The protocluster candidates are determined based on rich groups at $z \geq 2$ and their surrounding galaxies and groups. %We discuss the probability of these protocluster candidates growing into clusters. 
Our results are summarized as the following:    

\begin{itemize}

\item Most groups at $z \sim 5$ only contain galaxy pairs of similar luminosity. The large separation of galaxy pairs relative to the halo radius indicates that the galaxy or halo major mergers are not frequent at such a high redshift.   
    
\item At $z \sim 4$, the majority of most massive groups possess 3 members. Typically, a pair of galaxies is close to each other and has lower luminosity compared with the third member, which implies that the distinguish between BCG and FSG becomes to be significant. 
    
\item There are 400 groups at $z \sim 3$ with at least 5 member galaxies. The cases of groups shown have one or two prominent brightest galaxies, which may indicate that these galaxies have become the dominant BCGs by devouring their closest FSGs.
    
\item We use the ratio between the minor and major axis of the scatter ellipse to investigate the distribution of member galaxies or the shape of rich groups at $z \sim 3$. The scatter ellipse is calculated using the two coordinates RA and Dec of member galaxies. The groups at this early stage are more elongated than those with similar mass at low redshifts. This is consistent with the framework of LSS formation that galaxies distributed in clusters are more spherical than those in the filaments due to the earlier formations of filaments.
    
\item The groups at $z \sim 2$ become richer. There are 914 groups with a median mass of $1.8 \times 10^{13} \msunh$ possessing at least 10 members. As shown in their images, some galaxies with similar redshifts as the groups frequently appear around the boundary of the groups. These galaxies are supposed to contribute to the growth of groups at a later time.
    
\item Our groups/clusters at lower redshifts are matched well with the group sample from \citet{Oguri2017}, which is produced with the HSC-SSP PDR2 data set. Most matched groups in our sample have a halo mass larger than $10^{13.5} \msunh$. 
%\adr{We also compare with the group catalog from \citet{Yang2021}, who constructed millions of groups with DESI Legacy Imaging Surveys data set....}
    
\item We use the total number of neighboring galaxies, $N_{\rm nei}$, and the total halo mass of neighboring groups, $M_{\rm nei}$, to find high redshift protocluster candidates. These candidates are determined from the high redshift ($z \geq 2$) rich groups. We judge the probability of protocluster candidates to grow into clusters based on the criteria $N_{\rm cri}(z)$, which is defined as the number of remaining member galaxies in lower redshift massive groups at higher redshift according to the magnitude limit. We divide the probabilities into three levels: low, medium and high, based on the value of $N_{\rm nei}$ relative to $N_{\rm cri}(z)$. Most groups at $2 \leq z < 3$ with at least 10 member galaxies can be directly regarded as protocluster candidates according to their $M_{\rm nei}$.
\end{itemize}

While our groups and protocluster candidates catalog is produced mostly using photometric redshifts, it is interesting to note that some samples already have a number of spectroscopic and multi-wavelength observations. Besides, as the sky area used to study the science theme Galaxy Evolution in PFS is completely overlapped by the CLAUDS and HSC-SSP data set, once PFS starts operating, we will continuously update
our group catalogs using the massive spectroscopic data once they are available. 
Our catalog can be combined with the galaxies observation from PFS to investigate the properties and evolution of galaxies at high redshifts. Our group and protocluster candidate catalogs can be obtained from this link: \url{https://gax.sjtu.edu.cn/data/PFS.html}, or at Zenodo\footnote{https://doi.org/10.5281/zenodo.6516482}.

%Moreover, the properties of groups shown in the cases need to be further investigated with statistic analysis from observations and confirmation from simulations.

% Besides, as the follow-up of HSC observations, the PFS \citep{Takada2014} will obtain spectroscopic information of the sources selected from HSC. In the future, the performances of our catalog are expected to be better after updating accurate redshifts from PFS.

\acknowledgments
We sincerely thank the anonymous referee for the useful comments.
This work is supported by the national science foundation of China (Nos. 11833005, 11890691, 11890692, 11621303, 11890693, 11933003, 12173025), 111 project No. B20019, and Shanghai Natural Science Foundation, grant No. 19ZR1466800. We acknowledge the science research grants from the China Manned Space Project with Nos. CMS-CSST-2021-A02, CMS-CSST-2021-A03. Y.S.D. acknowledges the science research grants from the National Key R$\&$D Program of China via grant No. 2017YFA0402704, NSFC grants 11933003, and the China Manned Space Project with No. CMS-CSST-2021-A05. WC is supported by the STFC AGP Grant ST/V000594/1. He further acknowledges the science research grants from the China Manned Space Project with NO. CMS-CSST-2021-A01 and CMS-CSST-2021-B01. JH acknowledges the science research grants from NSFC grant 11973032, National Key Basic Research and Development Program of China (No.2018YFA0404504), and the sponsorship from Yangyang Development Fund.

These data were obtained and processed as part of the CFHT Large Area U-band Deep Survey (CLAUDS), which is a collaboration between astronomers from Canada, France, and China described in \citet{Sawicki2019}. CLAUDS is based on observations obtained with MegaPrime/ MegaCam, a joint project of CFHT and CEA/DAPNIA, at the CFHT which is operated by the National Research Council (NRC) of Canada, the Institut National des Science de l’Univers of the Centre National de la Recherche Scientifique (CNRS) of France, and the University of Hawaii. CLAUDS uses data obtained in part through the Telescope Access Program (TAP), which has been funded by the National Astronomical Observatories, Chinese Academy of Sciences, and the Special Fund for Astronomy from the Ministry of Finance of China. CLAUDS uses data products from TERAPIX and the Canadian Astronomy Data Centre (CADC) and was carried out using resources from Compute Canada and Canadian Advanced Network For Astrophysical Research (CANFAR).

The Hyper Suprime-Cam (HSC) collaboration includes the astronomical communities of Japan and Taiwan, and Princeton University. The HSC instrumentation and software were developed by the National Astronomical Observatory of Japan (NAOJ), the Kavli Institute for the Physics and Mathematics of the Universe (Kavli IPMU), the University of Tokyo, the High Energy Accelerator Research Organization (KEK), the Academia Sinica Institute for Astronomy and Astrophysics in Taiwan (ASIAA), and Princeton University. Funding was contributed by the FIRST program from the Japanese Cabinet Office, the Ministry of Education, Culture, Sports, Science and Technology (MEXT), the Japan Society for the Promotion of Science (JSPS), Japan Science and Technology Agency (JST), the Toray Science Foundation, NAOJ, Kavli IPMU, KEK, ASIAA, and Princeton University. 

This paper makes use of software developed for Vera C. Rubin Observatory. We thank the Rubin Observatory for making their code available as free software at http://pipelines.lsst.io/.

This paper is based on data collected at the Subaru Telescope and retrieved from the HSC data archive system, which is operated by the Subaru Telescope and Astronomy Data Center (ADC) at NAOJ. Data analysis was in part carried out with the cooperation of Center for Computational Astrophysics (CfCA), NAOJ.

This work has made use of the Gravity Supercomputer at the Department of Astronomy, Shanghai Jiao Tong University.





%\appendix


\bibliography{paper}{}
\bibliographystyle{aasjournal}



\end{document}